\documentclass[12pt,pra,notitlepage,tightenlines,reprint]{revtex4-1}

\usepackage{physics} 
\usepackage{bm} 
\usepackage{amsmath}
\usepackage{amssymb}
\usepackage{graphicx}
\usepackage[large]{subfigure}
\usepackage{color,enumerate}
\usepackage{soul,xcolor}

\newcommand{\rvec}{\mathbf{r}} 

\begin{document}

\setstcolor{red}
\title{Excited-State Trions in Two Dimensional Materials}

\author{Jun Yan}
\affiliation{Department of Physics, University of Massachusetts, Amherst,
Massachusetts 01003, USA}
\author{Kalman Varga}
\email{kalman.varga@vanderbilt.edu}
\affiliation{Department of Physics and Astronomy, Vanderbilt University, Nashville, Tennessee, 37235, USA}

\begin{abstract}
Using the  complex scaling and the stabilization method combined with
the stochastic variational approach, we have shown that there are narrow
resonance states in two-dimensional three particle systems  
of electrons and holes interacting via screened Coulomb interaction. 
These resonances are loosely bound systems of excited state excitons
with a third particle circling around them. Recent experimental
studies of excited state trions might be explained and identified by these
resonant states.
\end{abstract}
\maketitle

\section{Introduction}
Monolayers of transition metal dichalcogenides (TMD) are chemically
and mechanically stable making them ideal systems for studying physics
in two dimension (2D). The reduced dimensionality leads to a notably strong Coulomb 
interaction between charge carriers \cite{PhysRevLett.105.136805}.
This enhanced interaction, in turn, leads to the formation of tightly bound excitons
\cite{PhysRevLett.113.076802,Ugeda2014,dark_exciton,PhysRevLett.113.026803,
prl.120.057405}, charged excitons (trions) \cite{Mak2013,Zhu12082014,PhysRevB.90.075413,
PhysRevB.89.205436,PhysRevLett.123.167401}, and biexcitons
\cite{You2015,PSSR:PSSR201510224,doi:10.1021/acs.nanolett.8b03918,
Stevens2018,Barbone2018,Chen2018,Li2018,Ye2018,He2020,doi:10.1021/acs.jpcc.9b11582}.

Theoretical studies 
\cite{PhysRevB.92.205418,PhysRevB.88.045318,PhysRevB.89.125309,
PhysRevB.84.085406,PhysRevB.101.125423,PhysRevLett.114.107401,PhysRevB.92.161404,
PhysRevB.92.195305,doi:10.1021/acs.nanolett.5b03009,PhysRevB.93.125423,PhysRevB.91.245421,
PhysRevB.101.075302,PhysRevB.99.241301,PhysRevB.96.035131,PhysRevB.99.085301,PhysRevB.96.075431,
PhysRevB.95.035417,PhysRevB.100.115430}
played an important role in predicting the
stability and properties of these electron-hole complexes. Energies of excitons can be
calculated by solving the Bethe-Salpeter equation in the quasiparticle
band structure framework \cite{PhysRevB.84.085406,PhysRevB.101.125423,PhysRevB.96.201113}.
The effective mass approach with 2D interaction potential has also been
successfully used to calculate binding energies
\cite{PhysRevB.92.205418,PhysRevB.88.045318,PhysRevLett.114.107401,PhysRevB.92.161404,
PhysRevB.92.195305,doi:10.1021/acs.nanolett.5b03009,PhysRevB.93.125423,PhysRevB.91.245421,
Kezerashvili2019,PhysRevB.96.035131,PhysRevB.96.075431,PhysRevB.100.115430,PhysRevB.97.195408,PhysRevB.96.085302}
in good agreement with the BSE approach and the experimental results.
In the effective mass models the excitonic systems are considered to
be few particle systems, e.g. the trion is bound state of three
particles. Other interpretations also exist where trions are described as excitons
dressed by interactions with a Fermi sea of excess carriers \cite{PhysRevB.95.035417}.

Recent experimental studies have shown the existence of excited state trions
in TMDs \cite{PhysRevLett.123.167401,yan}. This is somewhat surprising, because the 
trion has no known bound excited state. In fact neither the H$^-$
(p,e$^-$,e$^-$) nor the Ps$^-$ (e$^+$,e$^-$,e$^-$) ion has bound
states in two or three dimensions \cite{appliedECG,PhysRevB.59.5652,varga99a,PhysRevB.61.13873}.
The H$^-$ and Ps$^-$ ion, however, has many resonant states in three dimensions 
\cite{PhysRevA.19.2347,PhysRevA.66.010502,Mezei2007}.

In this paper we will investigate the existence of these resonant
states in two dimensional materials. Unlike bound states,
the resonances have complex energies and spatially extended  non-L$^2$
wave functions. Conventional variational approaches based on square
integrable real basis functions can not be directly used to calculate
these resonant excited states. We will use two distinct approaches, the real 
stabilization method \cite{PhysRevA.1.1109}, and the complex scaling
(CS) \cite{HO19831,MOISEYEV1998212} approach to find the resonant states.
Both of these approaches need a flexible variational basis. We will
use the stochastic variational method (SVM) \cite{svmbook} with
explicitly correlated Gaussians (ECG) \cite{appliedECG} to generate basis states.

The stabilization method (SM) \cite{PhysRevA.1.1109} is based on the 
observation that a sufficiently large-square integrable basis set yields good 
approximations to the inner part of the exact resonance wave functions
at energies  equal to the eigenvalues of the 
Hamiltonian matrix. Eigenvalues belonging to resonant states remain
stable when the basis dimension is increased. The degree of stability 
of the eigenvalues approximating the energy of the resonance
is proportional to the width of the resonance. The complex energy
of the resonance state can be extracted from the change in
the stable eigenvalue as the size of the basis increases.

In the complex scaling method \cite{HO19831,MOISEYEV1998212}, the coordinates are rotated into the
the complex plane and resonant wave function
becomes square-integrable and can be expanded in terms of real basis
functions. Trajectory of the eigenenergies of the Hamiltonian as a
function of rotation angle are very different for bound, scattering and
resonance states. The energy and width of the resonances can be
determined from the converged position of the complex eigenvalues.

The stochastic variational method will be used to generate 
square-integrable basis using  explicitly correlated Gaussians
\cite{appliedECG} for the CS and SM calculations.  The SVM  
has been previously shown to achieve accuracy of up to 8-10
digits when describing the binding energies of similar systems such as
$\text{H}_2$, $\text{H}_2^+$, and the positronium molecule
($\text{Ps}_2$) \cite{varga99a,PhysRevA.58.1918}. This method has
proven to be
well-suited for describing the binding energies of excitonic structures ranging
from the two-body exciton to five-body exciton-trion systems
\cite{doi:10.1021/acs.nanolett.5b03009,PhysRevB.59.5652,PhysRevB.61.13873}. 
Previously, we have shown that this method yields values that 
agree with other calculations and experimental findings for the binding energies of excitons 
and trions in TMDs \cite{doi:10.1021/acs.nanolett.5b03009,PhysRevB.93.125423}.

\section{Formalism}
\label{sec:formalism}
\subsection{Hamiltonian and basis functions}

The nonrelativistic Hamiltonian of an excitonic few-particle system is given by
\begin{equation}
H=-\sum_{i=1}^N  \frac{\hbar^2}{2m_i} \nabla^2_i
+\sum_{i<j}^N V(r_{ij}),
\label{ham}
\end{equation}
where $r_{ij}=|\mathbf{r}_i-\mathbf{r}_j|$, and $\mathbf{r}_i$, $m_i$, are the 2D position vector and the effective mass
of the particle.

In the case of an excitonic system in  2D, the interaction 
potential $V(r_{ij})$ is given by the 2D screened electrostatic interaction
potential derived by Keldysh \cite{Keldysh}
\begin{equation}
\label{eq:Keldysh}
V(r_{ij})=\frac{q_i q_j}{\kappa r_0}V_\text{2D}\left ( \frac{r_{ij}}{r_0} \right ),
\end{equation}
where
\begin{equation}
V_\text{2D}(r)=\frac{\pi}{2} \left [ H_0(r) - Y_0(r) \right ].
\end{equation}
This potential has been adopted in most of the calculations. Alternative
potentials have also been proposed \cite{PhysRevB.98.125308} to better
describe three atomic sheets that compose a monolayer TMD. 
In the screened potential $q_i$ is charge of the $i$th particle,
and $r_0$ is the screening length indicative of the medium. $\kappa$ is the average 
environmental dielectric constant. $H_0$ and $Y_0$ are the 
Struve function and Bessel function of the second kind, respectively. 

The nonlocal macroscopic screening, inherent to 2D systems, distinguishes 
this potential from its 3D Coulombic counterpart
\cite{PhysRevB.84.085406}. The length scale of 
this screening is determined by the 2D layer polarizability
$\chi_\text{2D}$ as $r_0=2\pi\chi_\text{2D}/\kappa$. In the limit of very strong 
screening ($r_0\rightarrow \infty$), 
the potential exhibits a logarithmic divergence, while in the limit of small
screening length ($r_0\rightarrow 0$), $V(r_{ij})$ approaches the 
usual $1/r$ behavior of the Coulomb potential.

The variational method is used to calculate the energy of the
system. As a trial function we choose a two dimensional (2D) form of the correlated Gaussians 
\cite{svmbook,appliedECG}:
\begin{equation}
\exp{ -\frac{1}{2} \sum_{i,j=1}^N  A_{ij} \mathbf{r}_i \cdot \mathbf{r}_j },
\end{equation}
where $A_{ij}$ are the nonlinear parameters. The above form of the CG belongs to $M=0$. To allow for 
$M\ne0$ states, we multiply the basis by
\begin{equation}
\prod_{i=1}^N \xi_{m_i}(\mathbf{r}_i) ,
\end{equation}
where
\begin{equation}
\xi_m(\bm{\rho}) = (x + iy)^m .
\end{equation}
Thus our nonrestrictive CG function reads as
\begin{multline}
\label{eq:trial}
\Phi_A(\rvec) = \mathcal{A} \left\{ \qty( \prod_{i=1}^N \xi_{m_i}(\mathbf{r}_i) ) \right. \times \\
 \left. \exp{ -\frac{1}{2} \sum_{i,j=1}^N A_{ij} \mathbf{r}_i \cdot \mathbf{r}_j} \right\},
\end{multline}
where $M = m_1 + m_2 + \cdots + m_N$, $m_i$ are integers, and $\mathcal{A}$ 
is an antisymmetrizing operator. This function is
coupled with the spin function $\chi_{SM_S}$ to form the trial
function. The nonlinear parameters are optimized using the stochastic
variational method \cite{appliedECG,svmbook}.

Explicitly Correlated Gaussians  are very popular in atomic
physics and quantum chemistry \cite{appliedECG}.
The main advantages of ECG bases are: (1) their matrix elements are
analytically available for a general N-particle system, (2) they are
flexible enough to approximate rapidly changing functions, (3) the permutation
symmetry can be easily imposed.

Ref. \cite{appliedECG} provides a thorough review of the
applications of the ECG basis in various problems. Benchmark tests
presented for atoms with $N$=2--5 electrons show that the ECG basis can 
produce up to 10 digit accuracy for 2--3 electron atoms. The ECG basis has also
proven to be very accurate in calculating weakly bound states. A
series of positronic atoms have been predicted using the stochastic
variational method with an ECG basis
\cite{PhysRevA.61.062503,varga99a}.
The binding energy of these systems \cite{appliedECG}
ranges from 0.001 to 0.04 a.u. (1 a.u. is 27.211 eV) with weakly bound diffuse structures 
similar to those studied here. 

\subsection{Complex scaling}
The complex scaling method was originally proposed by Aguilar, Balslev, and Combes
\cite{aguilar71,balslev71}.
The CS is introduced by a transformation $U(\theta)$ with a scaling angle
$\theta$ for the radial coordinate $\rvec$ 
\begin{equation}
U(\theta)\, \rvec\, U^{-1}(\theta)=  \rvec\ e^{i\theta}, 
\end{equation}
where $U(\theta) U^{-1}(\theta)=1$.
The Schr\"odinger equation, $H\Psi=E\Psi$, is transformed as
\begin{eqnarray}
H^\theta \Psi^\theta
&=&    E^\theta \Psi^\theta,\qquad \\
H^\theta &=&   U(\theta)HU^{-1}(\theta),
\label{cse}
\end{eqnarray}
To solve  Eq. \ref{cse}, the wave functions $\Psi^\theta_k(\rvec)$ are expanded in terms of
ECG basis functions:
\begin{equation}
      \Psi^\theta_k(\rvec)=\sum_{i=1}^{K} c_{ik}(\theta)\, \Phi_{A_i}(\rvec),
\end{equation}
leading to the generalized complex eigenvalue problem
\begin{eqnarray}
      \sum_{j=1}^{K}H_{ij}^\theta\, c_{jk}(\theta)
&=&   E^\theta_k\, \sum_{j=1}^K O_{ij} c_{jk}(\theta),
      \\
    H_{ij}^\theta
&=& \langle \Phi_{A_i} | H^\theta | \Phi_{A_j} \rangle \\
    O_{ij}
&=& \langle \Phi_{A_i} | \Phi_{A_j} \rangle ,
\end{eqnarray}
where $H_{ij}^\theta$ are the matrix elements of the complex-scaled Hamiltonian 
and $O_{ij}$ is the overlap of the basis functions. In the case of
Coulomb interactions the CS Hamiltonian is particularly simple:
\begin{equation}
H_{ij}^\theta={\rm e}^{-2\theta} T_{ij} + {\rm e}^{-\theta} V_{ij} 
\end{equation}
where $T_{ij}$ and $V_{ij}$ are the kinetic and potential energy
matrices of  the ECG basis functions.

The ABC theorem of Aguilar, Combes, and
Balslev\cite{aguilar71,balslev71} describes the properties of the 
of the CS eigenstates:

\begin{enumerate}[(a)]
\itemsep=0.1cm
\item Energies of  bound states are invariant with respect to
the rotation angle.
\item Resonance states can be described by  square-integrable functions.
\item The continuum spectra start at the threshold energies
corresponding to the decays of the system into subsystems. The spectra
is rotated clockwise by  $2\theta$ from the positive real energy axis.
\end{enumerate}

In the CS method the resonances are determined by finding the position
where the complex eigenvalues are stabilized  with respect
to the rotation angle:
\begin{equation}
\left. {\partial E\over\partial \theta}\right\vert_{\theta=\theta_{opt}}=min.
\end{equation}
Once the position of resonance  is determined, the resonance
energy ($E_r$) and total width ($\Gamma$) are given by
\begin{equation}
E= E_r - i {1\over 2} \Gamma.
\end{equation}

\subsection{Stabilization method}
In this method we also use a variational ECG basis ansatz
\begin{equation}
      \Psi^{(K)}_k(\rvec)=\sum_{i=1}^{K} c_{ik}\, \Phi_{A_i}(\rvec),
\end{equation}
but now we use the dimension of the basis, $K$, as a parameter
(expansion length). The corresponding  generalized real eigenvalue
problem reads as
\begin{eqnarray}
      \sum_{j=1}^{K}H_{kj}, c_{ji}(\theta)
&=&   \epsilon^{(K)}_i\, \sum_{j=1}^K O_{kj} c_{ji}(\theta),
      \\
    H_{ij}^\theta
&=& \langle \Phi_{A_i} | H| \Phi_{A_j} \rangle, \\
\end{eqnarray}
where $\epsilon^{(K)}_i$ is the variational estimate to the energy of
the $i$th state of the system.

The simplest version of the stabilization method \cite{PhysRevA.1.1109} is
based on the Hylleraas-Undheim theorem \cite{Hylleraas1930,PhysRev.43.830}.
The Hylleraas-Undheim theorem states that (i) comparing the variational
energy estimates obtained with $K$ trial wavefunctions and
the estimate obtained by adding one additional orthonormalized trial wavefunction
(increasing the basis dimension to $K+1$), one finds
that the new energy estimates are interleaved with the old ones:
\begin{equation}
\epsilon_0^{(K+1)}\le
\epsilon_0^{(K)}\le
\epsilon_1^{(K+1)}\le
\epsilon_1^{(K)}\le
{\ldots} 
\epsilon_{K-1}^{(K+1)}\le
\epsilon_{K-1}^{(K)}\le
\epsilon_K^{(K+1)}
\end{equation} 
and (ii) the eigenvalues $\epsilon^{(K)}_i$ are upper limits to the 
corresponding excited states. 

By increasing the basis dimension (``expansion length'') 
the real part of the resonance energies become  ``stable''.
This stabilization is due to the fact that the inner part
of the wave function, at an energy in the resonant region, 
looks like the wave function of a bound state. 
The amplitude of the  wavefunction in the asymptotic region is
much smaller than the amplitude of its inner part. 
The inner part of the wave function is expanded in a
set of discrete exponentially decaying functions,
and then the Hamiltonian is diagonalized to yield
the approximate resonance energies directly. Once the
basis size is sufficiently large to represent the inner
part, the energy of this state barely changes when more basis
states are added, because the asymptotic part is small and does
not contribute to the energy. The energies of the nonresonant 
scattering wave functions, however, quickly change with the addition
of basis states because their asymptotic parts are large.

There are many variants of the stabilization method, one can confine
the wave functions with a potential and change the range of
confinement, scale the coordinates 
or perturb the Hamiltonian in some way and find the stable states
\cite{Pont_2011,SAJEEV2013105,THOMPSON198271,MACIAS1989359,doi:10.1002/qua.560270414}. 
Most of these approaches can be used to extract the resonance widths
as well  \cite{Pont_2011,SAJEEV2013105,THOMPSON198271,MACIAS1989359,doi:10.1002/qua.560270414}.

\subsection{Stochastic optimization}
The basis parameters can be efficiently chosen via the stochastic 
variational method \cite{svmbook}. In this
approach, the variational parameters $A_{ij}$ of the ECG basis (see Eq.
\ref{eq:trial}) are randomly selected, and the parameters giving the 
lowest variational energy are retained as basis states. This procedure
can be fine-tuned into an efficient optimization scheme as described in
detail in Refs. \cite{appliedECG,svmbook}. 

In the present work we found that the most efficient way to build a
flexible basis is as follows>

(i) optimize the ground state on a small ($K=200$) basis selecting
$A_{ij}$ from  a parameter space which confines the interparticle
distances   below 10 a.u.

(ii) expand the basis by optimizing the lowest $l$ states by minimizing
\begin{equation}
\sum_{i=1}^l (\epsilon_i^{(K)}-\epsilon)^2
\end{equation}
for each $K$ by SVM. Here $\epsilon$ can be any number below the
lowest eigenvalue. In this step $A_{ij}$ is  chosen  allowing the interparticle
distances   to extend up to 100 a.u. This helps the description of the
extended excited states.

\subsection{Physical quantities}
The following physical quantities will be used to
describe the properties of the system and characterize the
quality of the wave function. The pair correlation function 
is defined as
\begin{equation} \label{corr_func_pair}
	C_{pq}(\rvec) = \frac{2}{N(N-1)} \ev**{ \sum_{i<j}^N
	\delta(\rvec_i-\rvec_j-\rvec) }{\Psi},
\end{equation}
where $p$ and $q$ stand for electrons or holes.
Using $C_{pq}(r)$, the radial part of the correlation function, the
powers of inter-particle distances are given by
\begin{equation}
\label{distances}
\ev{r^k_{pq}} = 2\pi \int_0^\infty r^k C_{pq}(r) r \dd{r}.
\end{equation}

\subsection{Units}
The effective electron and hole masses are denoted as
\begin{equation}
m_e^*=m_e m_0 \ \ \ \ \textrm{and}  \ \ \ \ m_h^*=m_h m_0
\end{equation}
where $m_0$ is the mass of the electron. One can define an effective Bohr radius as
\begin{equation}
a^*={\hbar^2\kappa\over \mu e^2}
\end{equation}
where $e$ is the electron's charge and 
\begin{equation}
\mu={m_e^*m_h^*\over m_e^*+m_h^*}
\end{equation}
is the reduced mass. This can be also written as
\begin{equation}
a^*={\kappa(1+\sigma)\over m_e}{\hbar^2\over m_0
e^2}={\kappa(1+\sigma)\over m_e} a_0 \ \ \
\ \sigma={m_e\over m_h}
\end{equation}
where $a_0={\hbar^2\over m_0 e^2}$ is the hydrogenic Bohr radius ($a_0=5.29177$ \AA). With
this one can define 
\begin{equation}
\tilde{r}_0 = r_0/a^*,
\end{equation}
the screening length normalized by the Bohr radius. Similarly, the
effective Bohr energy
\begin{equation}
E^*={e^2\over\kappa a^*}={m_e\over \kappa^2(1+\sigma)}E_0,
\end{equation}
where 
\begin{equation}
E_0={e^2/a_0}
\end{equation}
is the Hartree energy ($E_0=27.211$ eV).

The energy of the exciton and trion states only depends
on $\tilde{r}_0$ and on the electron-hole mass ratio $\sigma$. In the 
following we will use atomic units. The energies of the exciton is
analytically known for $\tilde{r}_0$=0 and $\sigma=1$ \cite{PhysRevA.43.1186}:
\begin{equation}
E_{ns}^X=-{2\over (2n-1)^2},
\end{equation}
that is -2, -2/9 and -2/25 a.u. for the $1s$,$2s$ and the $3s$ states
respectively.

To convert the results to eV and \AA \ one has to multiply the energies 
by $E^*$ and the distances by $a^*$.  This is the same convention as
used in Refs. \cite{PhysRevB.92.195305,PhysRevB.88.045318,PhysRevB.93.125423}

\vskip 1.cm
\section{Results and Discussion}

\subsection{Spin singlet case}
\begin{figure*}[htp]
\centering
\includegraphics[width=0.45\linewidth]{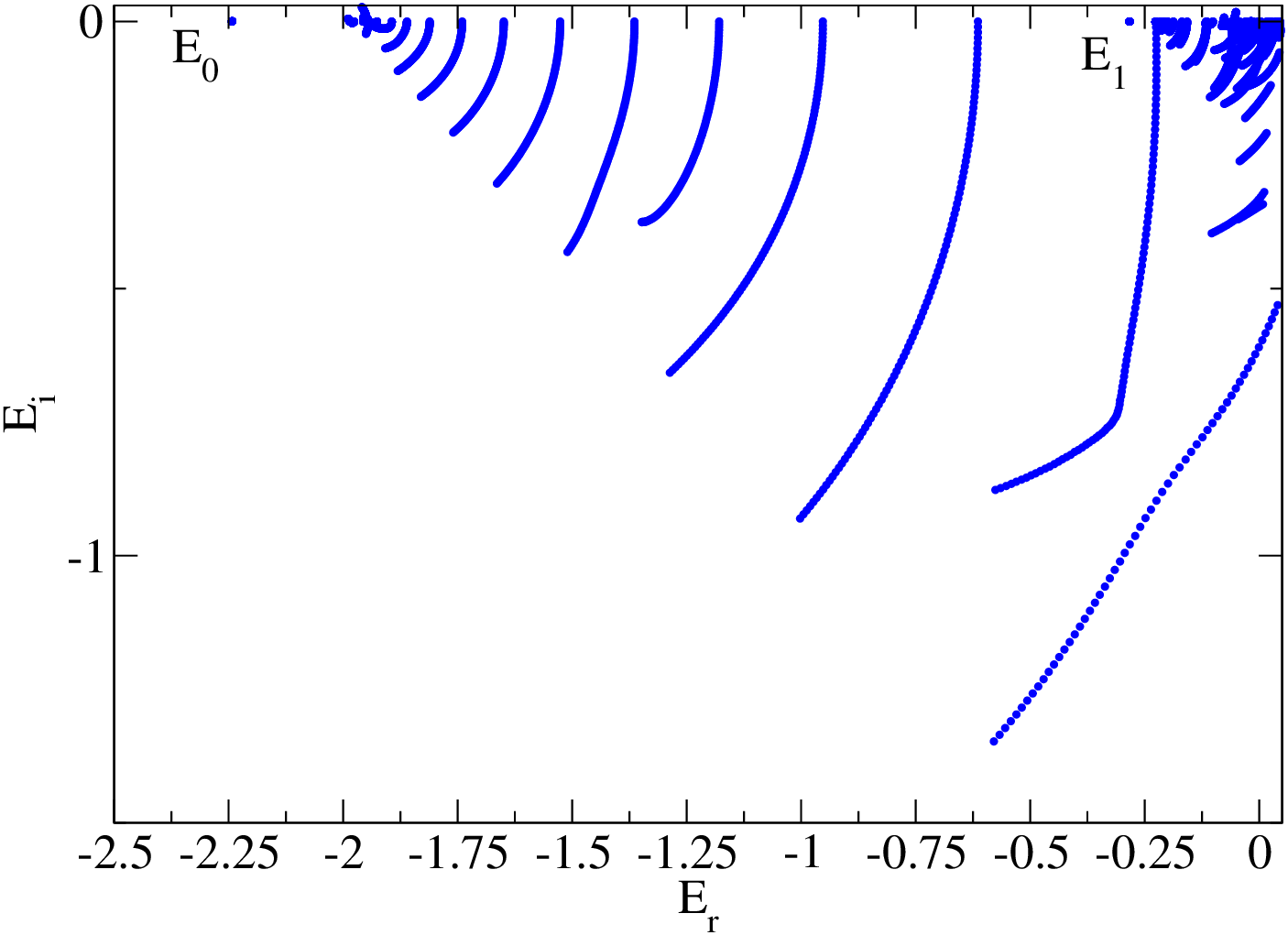} 
\caption{Trajectories of the complex rotated energies.
The energies are rotated from $\theta=0$ to $\theta=0.36$ (in rad) with a
stepsize of $\Delta\theta=0.004$ (in rad). $r_0=0$ is used.}
\label{fig1}
\end{figure*}

First we present the results for the spin singlet state of the trion.
As an illustration of the CS calculation, Fig. \ref{fig1}  shows the
lowest 50 eigenvalues as the function of the rotation angle for a trion
with $\sigma=0$ and $r_0=0$. The basis dimension in all calculations is $K=1500$.
The ground state energy, $E_0$, is below the $1s$ exciton energy
($E_{1s}^X=-2$ a.u.), and the resonance state ($E_1=-0.28$ a.u.) below the $2s$ exciton
threshold ($E_{2s}^X=-2/9$ a.u.) remains stable.  The  continuum
states rotated to the complex plane with an angle of 2$\theta$. The
first set of continuum states is rotated from the $1s$ threshold, the
second starts at the $2s$ threshold. 

\begin{figure*}[htp]
\centering
\includegraphics[width=0.45\linewidth]{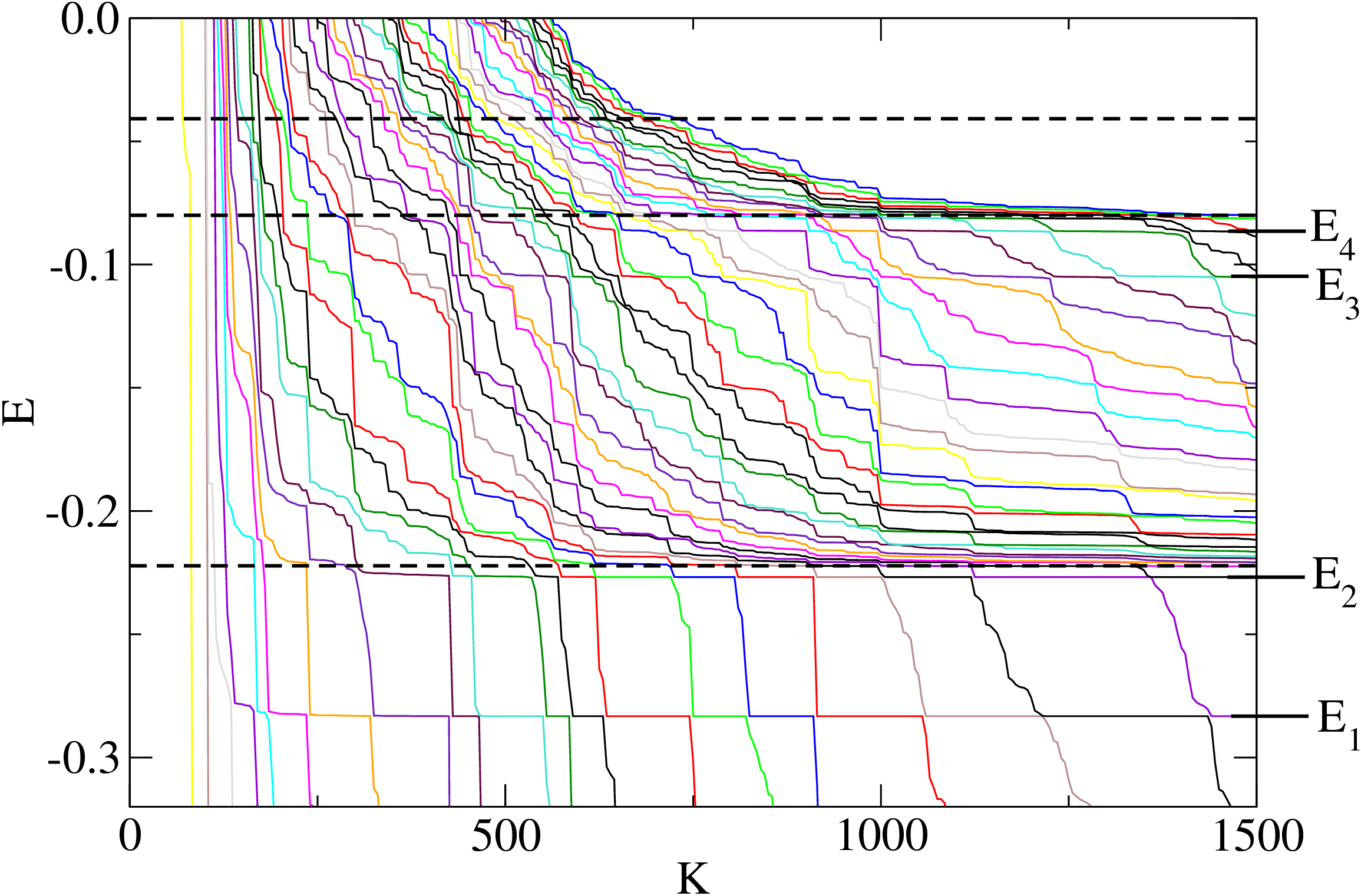} 
\caption{Energy levels versus basis dimension. The four excited states
energies ($E_1,E_2,E_3$ and $E_4$) are nicely stabilized as the energies
are converging with the increased basis size. $\tilde{r}_0$=0 is used. The
dashed lines show the $2s$, $3s$ and $4s$ exciton energies.}
\label{fig2}
\end{figure*}

Fig. \ref{fig2} show the stabilization of the of the energy of trion
resonance states. The lowest state, $E_1=$-0.28 a.u. (same state is
shown on Fig. \ref{fig1}) is stabilized at a small ($K$=200) basis
dimension and its energy remains unchanged after that. One can also see 
that at certain $K$ values there are ``avoided crossings'' of the
neighboring eigenvalues. The avoided crossing is a simple consequence of
the Hylleraas-Undheim theorem. Let's assume that an isolated ``stable'' eigenvalue 
$\epsilon^{(K)}_j$ approximates $E_1$. With increasing $K$ the next higher eigenvalue,
$\epsilon^{(K)}_{j+1}$, decreases more rapidly than the stable
eigenvalue $\epsilon^{(K)}_j$, and pushes $\epsilon^{(K)}_j$ away from 
$E_1$. At this point $\epsilon^{(K)}_{j+1}$ approaches $E_1$ and becomes
the stable eigenvalue.

\begin{widetext}
\begin{table*}
\begin{tabular}{|c|l|l|l|l|l|l|l|l|}
\hline
$\tilde{r}_0$&$E_r$   & $E_i$               & $r_{eh}$&$r_{eh}^2$&$\delta_{eh}$         & $r_{ee}$&$r_{ee}^2$&$\delta_{ee}$   \\\hline
     &-2.2432 &     0               &0.84 &1.20    & 9.76               & 1.30& 2.28   & 0.48                \\\cline{2-9}
     &-0.2832 &-8.8$\times 10^{-8}$ &4.09 &20.93   & 0.22               & 7.38& 59.98  & 9.52$\times 10^{-5}$ \\\cline{2-9}
     &-0.2267 &-1.4$\times 10^{-5}$ &9.98 &171.58  & 0.26               &18.98& 414.41 & 2.37$\times 10^{-5}$ \\\cline{2-9}
  0  &-0.1049 &-1.7$\times 10^{-4}$ &12.98&414.39  & 6.47$\times 10^{-2}$&24.07& 957.33 & 1.92$\times 10^{-4}$ \\\cline{2-9}
     &-0.0886 &-1.2$\times 10^{-4}$ &16.88&421.15  & 3.90$\times 10^{-2}$&32.10& 1179.79& 4.52$\times 10^{-4}$ \\\cline{2-9}
     &-0.0814 &-9.3$\times 10^{-5}$ &29.82&1463.13 & 2.86$\times 10^{-2}$&57.80& 3627.17& 8.11$\times 10^{-7}$ \\\hline
\hline
     &-0.8369 &     0               &1.70 & 4.72   & 1.51                &2.68  & 9.32    & 0.117                \\\cline{2-9}
     &-0.2217 &-3.4$\times 10^{-7}$ &5.42 & 42.86  & 7.36$\times 10^{-2}$&9.73 & 113.38  & 1.03$\times 10^{-4}$  \\\cline{2-9}
0.5  &-0.1926 &-5.8$\times 10^{-5}$ &29.06& 1954.90& 3.76$\times 10^{-2}$&55.01 & 3911.25 & 2.19$\times 10^{-3}$  \\\cline{2-9}
     &-0.0914 &-1.6$\times 10^{-4}$ &13.26& 287.85 & 4.50$\times 10^{-2}$&24.17 & 720.18  & 1.25$\times 10^{-3}$ \\\cline{2-9}
     &-0.0781 &-2.1$\times 10^{-4}$ &27.74& 1065.93& 1.65$\times 10^{-2}$&43.59 &2224.40  & 5.70$\times 10^{-5}$\\\hline
\hline
     &-0.5956 &      0              &2.14 &  7.39  & 0.90                & 3.39& 14.75  & 7.24$\times 10^{-2}$\\\cline{2-9}
     &-0.1918 &-2.2$\times 10^{-7}$ &6.19 & 51.91  & 5.56$\times 10^{-2}$&11.08& 138.89 & 5.29$\times 10^{-2}$\\\cline{2-9}
1    &-0.1715 &-6.7$\times 10^{-5}$ &29.85& 1960.28& 5.23$\times 10^{-3}$&56.23&3922.15 &3.84$\times 10^{-2}$\\\cline{2-9}
     &-0.0841 &-4.2$\times 10^{-4}$ &18.16&829.50  & 4.34$\times 10^{-2}$&32.73&1809.85 &1.93$\times 10^{-3}$ \\\cline{2-9}
     &-0.0740 &-3.9$\times 10^{-4}$ &25.11& 1049.68& 5.24$\times 10^{-3}$&45.55&2351.17 &8.46$\times 10^{-5}$ \\\hline
\end{tabular}
\caption{Energies and other properties of trion states for $S=0$. For $\tilde{r}_0$=0 the ground state and 5 resonance states is shown.
For $\tilde{r}_0=0.5$ and $\tilde{r}_0=1$ only four resonance state is listed because the energy of the  5th state is too close to the energy
of the $3s$ exciton and the accurate calculation is very difficult.}
\end{table*}
\end{widetext}

The next resonance state, $E_2$ is slightly below the $2s$ threshold
and it is also stable from about $K=300$. There are at least two more
stable resonant states, $E_3$ and $E_4$ below the $3s$ threshold. Fig.
\ref{fig2} also shows that many states converges to the $2s$ and $3s$
thresholds from above. These states represents dissociation into $2s$
and $3s$ excitons and an electron.

All these resonance states are stabilized as a straight vertical line
which is a typical sign of a narrow resonance. This is conformed by
complex scaling as we will see later.

Ref. \cite{PhysRevA.66.010502} also found 4 resonant states in the 
three dimensional case. The energies are obviously different in 2D and
3D, but there 
are two resonances below the $2s$ threshold in both 2D and 3D. 
Ref. \cite{PhysRevA.66.010502} shows two resonances below the $3s$
threshold, our complex scaling approach shows at least 3 states. 

\begin{figure*}
\subfigure{
\includegraphics[width=0.45\linewidth,clip]{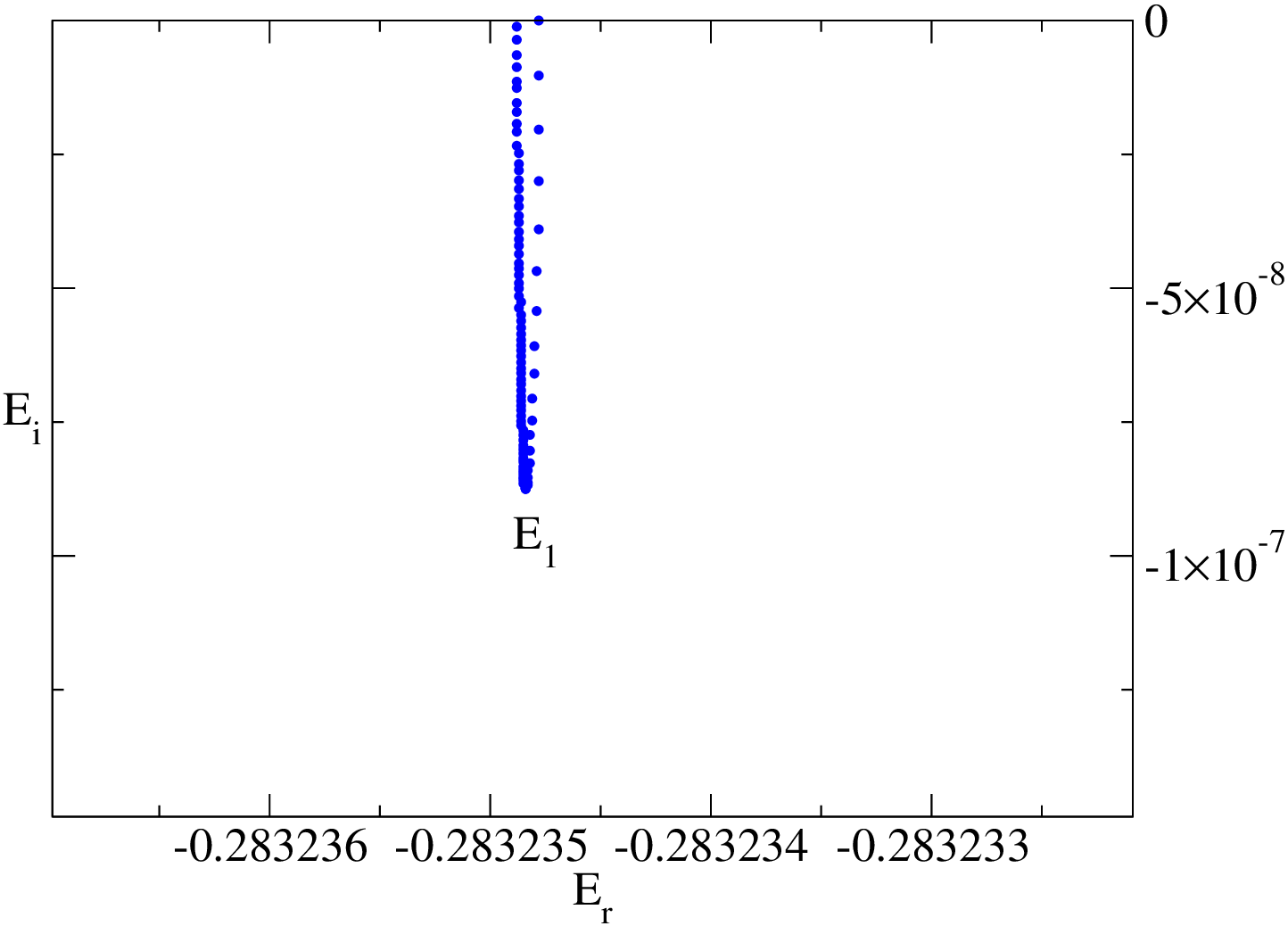} 
}
\subfigure{
\includegraphics[width=0.45\linewidth,clip]{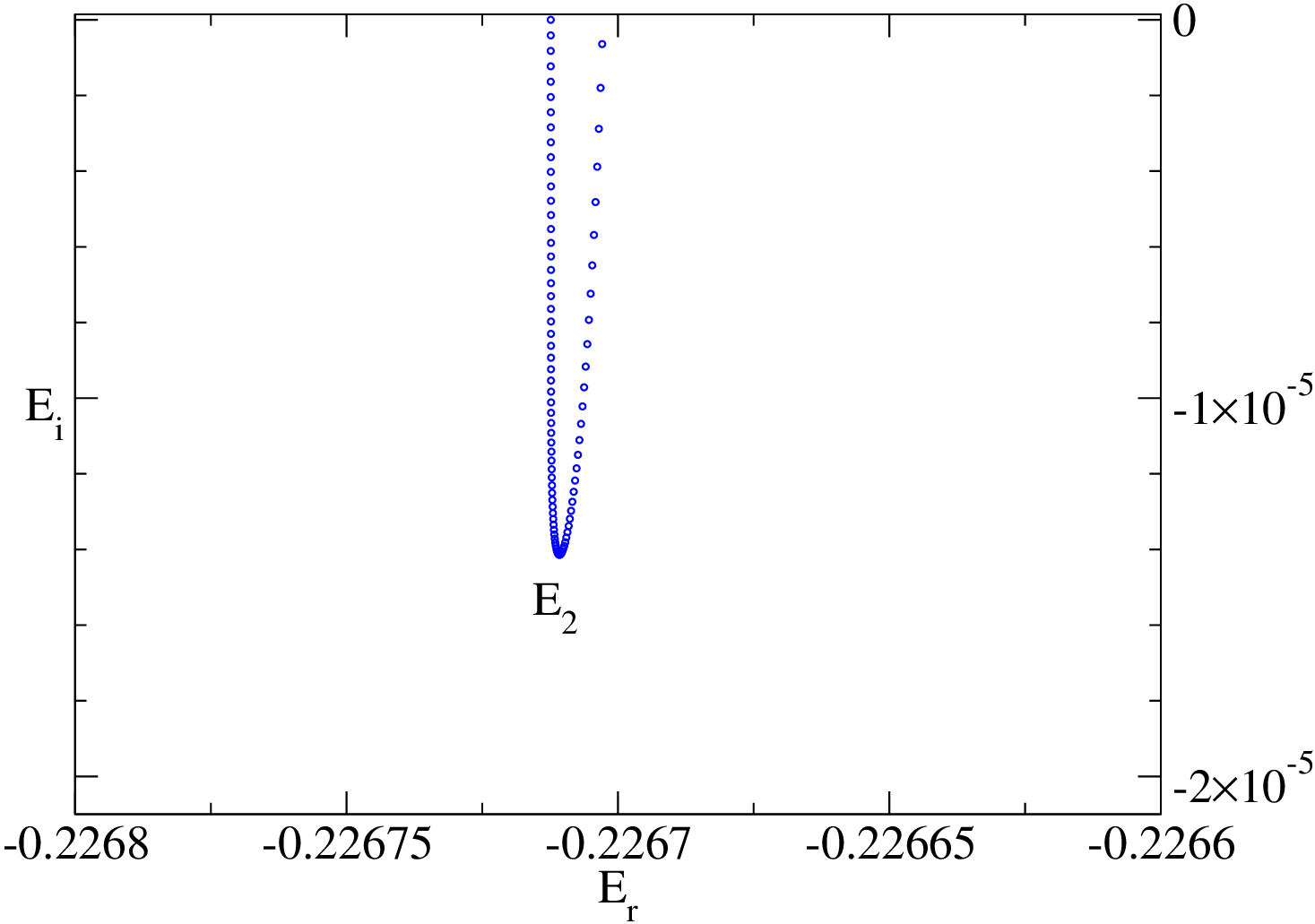}
}
 \\ 
\subfigure{
\includegraphics[width=0.45\linewidth,clip]{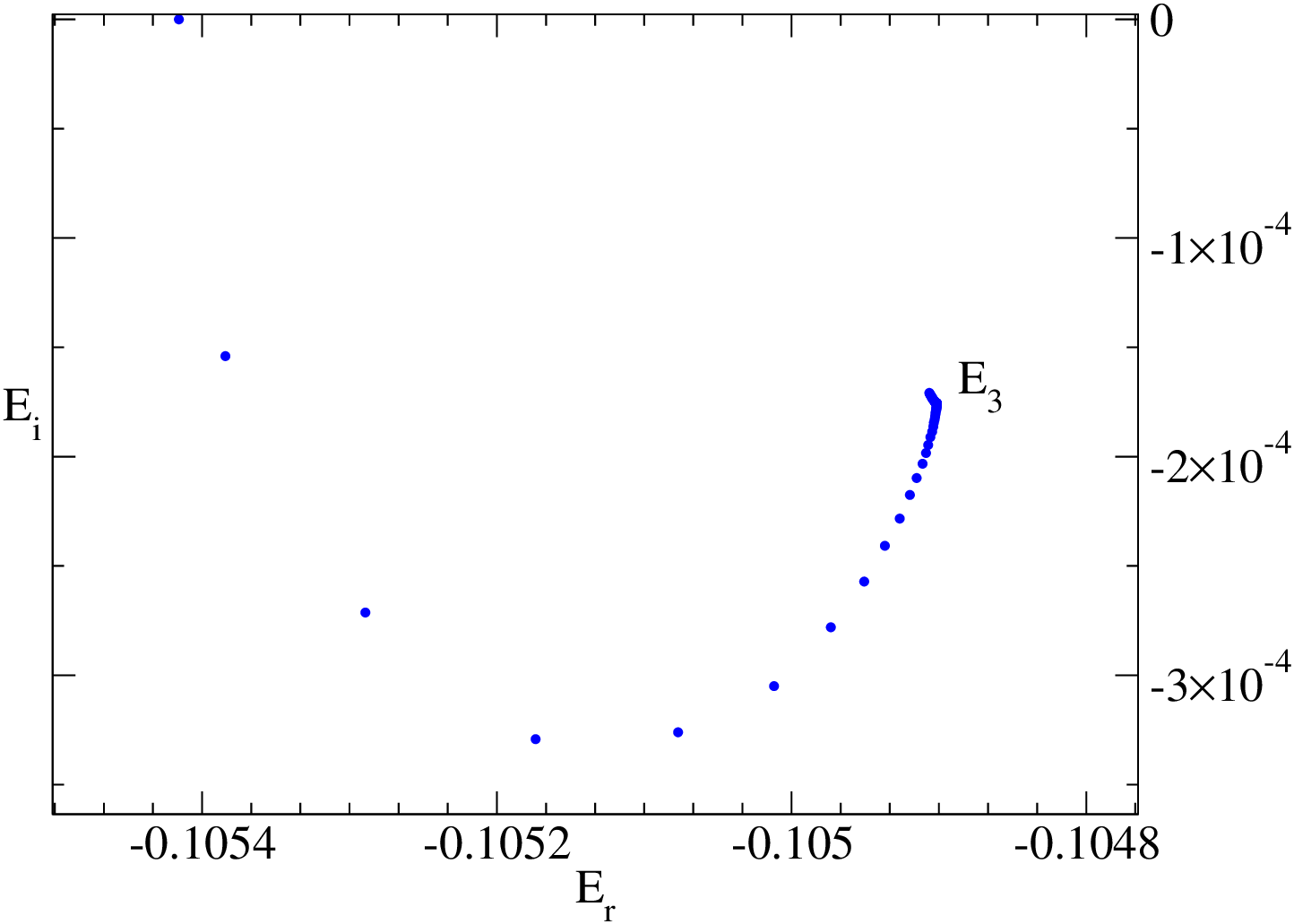} 
}
\subfigure{
\includegraphics[width=0.45\linewidth,clip]{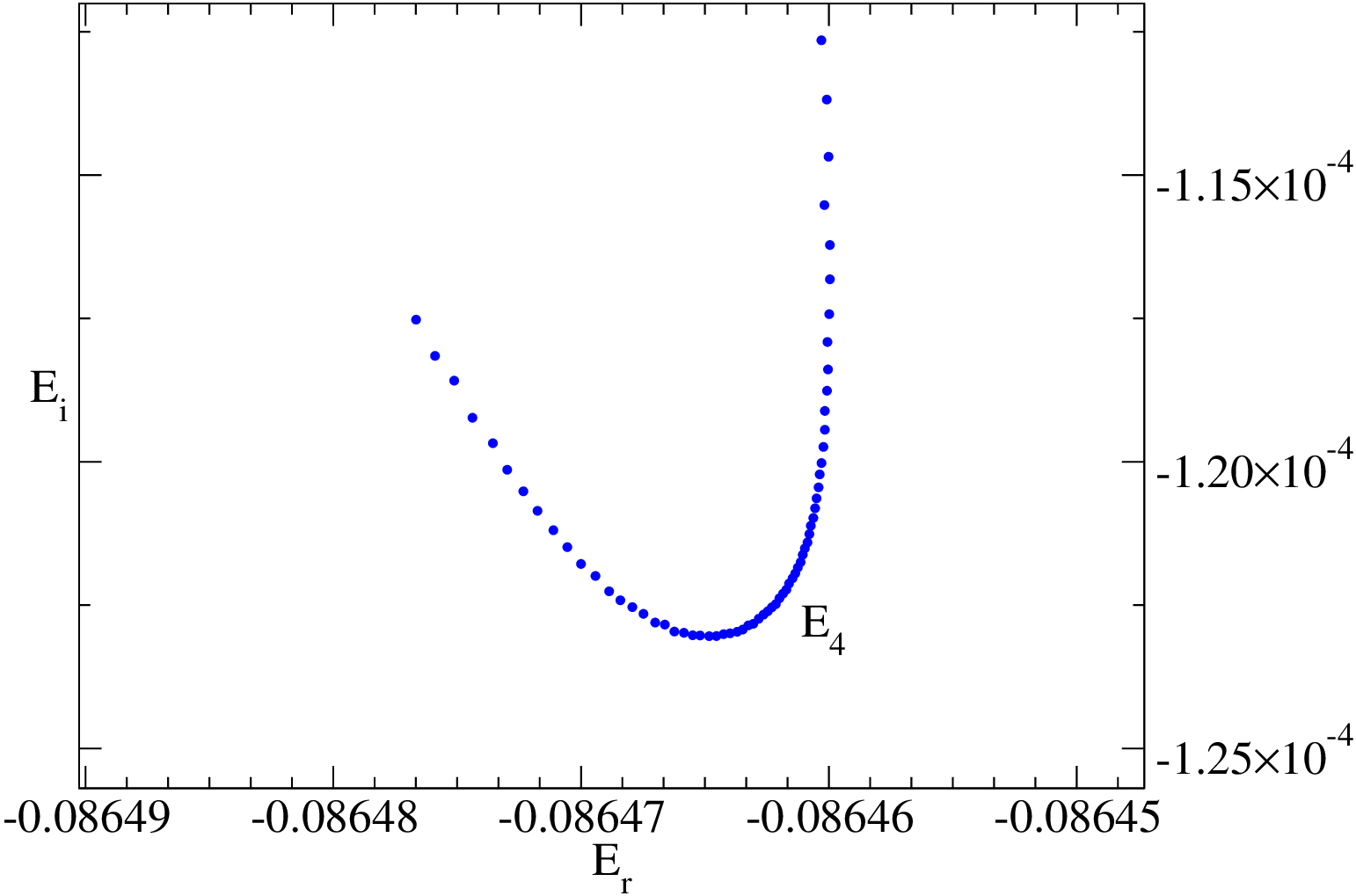}
}
\caption{Trajectories of the excited states of trion on the complex
energy plane. The energies are rotated from $\theta=0$ to $\theta=0.1$ with a
stepsize of $\Delta\theta=0.004$ (in rad). $\tilde{r}_0=0$ is used.}
\label{fig3}
\end{figure*}

By enlarging the CS calculation results in Fig. \ref{fig1} around the 
resonant states, Fig. \ref{fig3} shows 4 narrow resonance states in the complex energy
plane. This is  the same 4 states that is calculated by
the stabilization  method and shown in Fig. \ref{fig2}.  These are all
very narrow resonances.

\begin{figure*}[htp]
\centering
\includegraphics[width=0.95\linewidth]{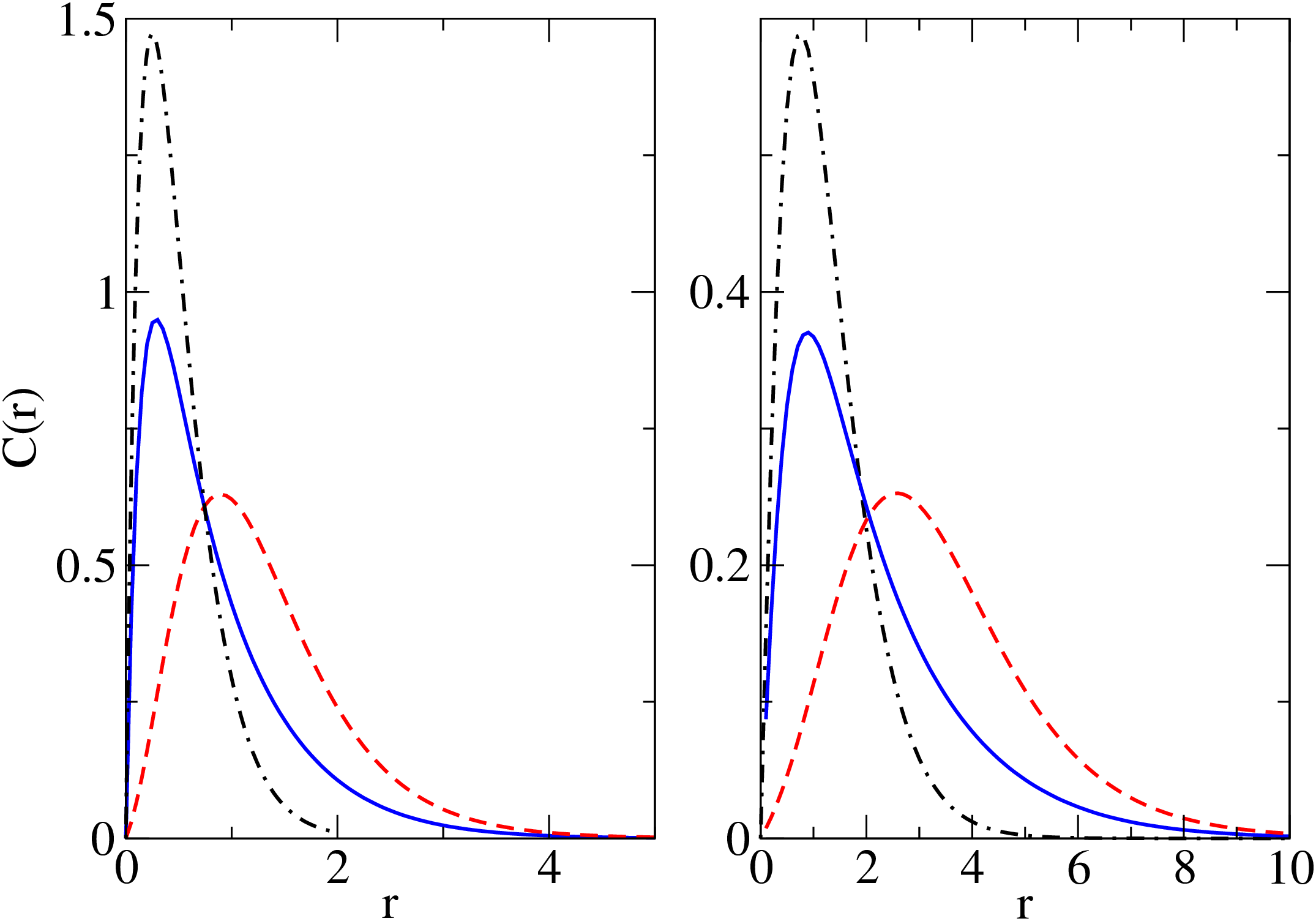} 
\caption{Ground state correlation functions for
$\tilde{r}_0=0$ (left) and $\tilde{r}_0=1$ (right).
Electron-electron correlation, $C^t_{ee}(r)$ (dashed line),
electron-hole correlation, $C^t_{eh}(r)$ (solid line), and 
exciton electron-hole correlation, $C^x_{ee}(r)$ (dashed-dotted line).}
\label{fig4}
\end{figure*}

Fig. \ref{fig4} shows the electron-electron and the electron hole
correlation functions for the ground state of trion. Due to the presence 
of the second electron the electron-hole correlation function is somewhat
wider than that of the exciton. The electron-electron correlation
function is pushed away from the origin due to the repulsion. The
structure of the trion is more or less similar to a system where an
electron is orbiting around an exciton. 

\begin{figure*}[htp]
\centering
\includegraphics[width=0.95\linewidth]{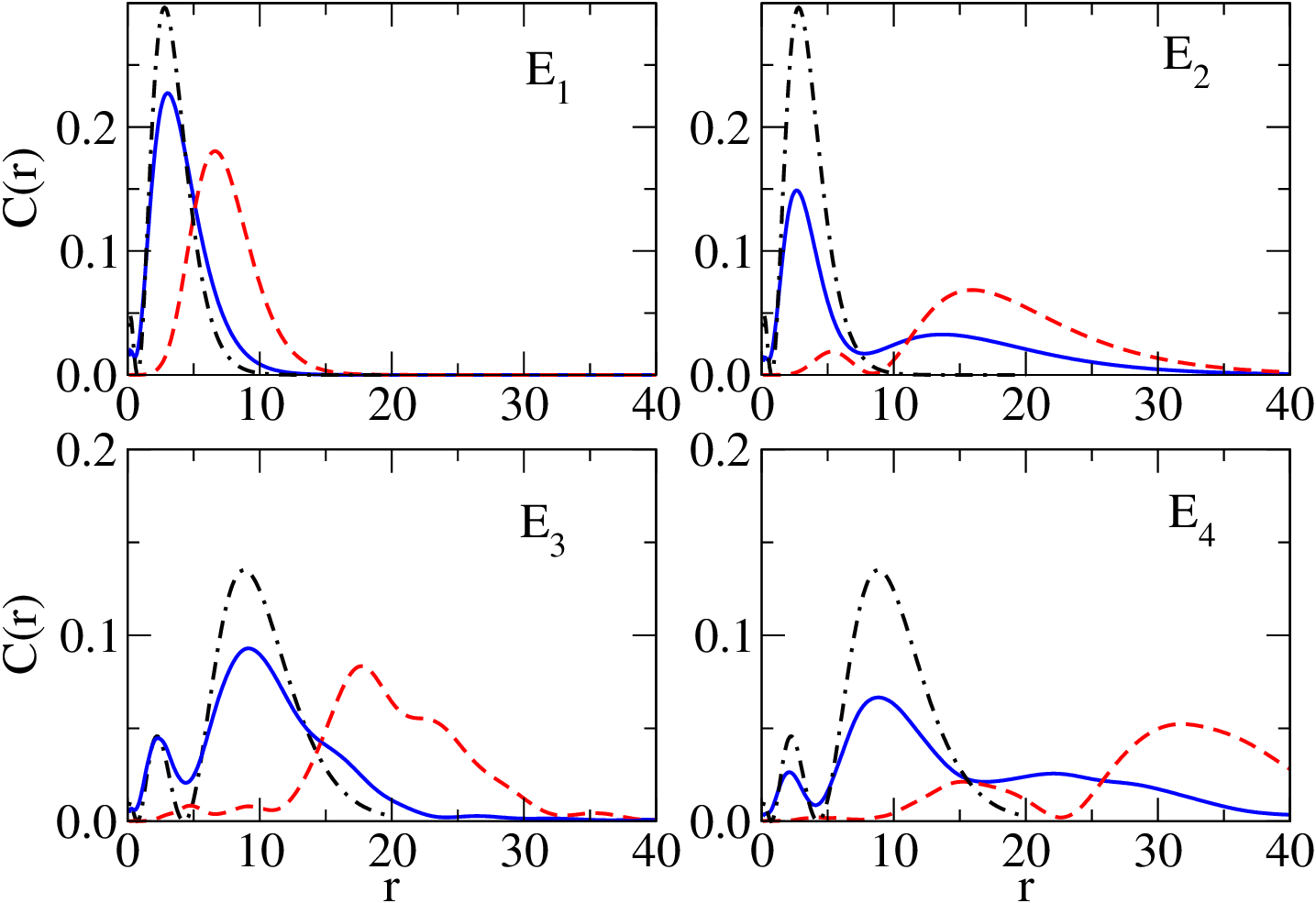} 
\caption{Excited state correlation functions for $\tilde{r}_0=0$.
Electron-electron correlation, $C^t_{ee}(r)$ (dashed line),
electron-hole correlation, $C^t_{eh}(r)$ (solid line), and 
exciton electron-hole correlation, $C^x_{ee}(r)$ (dashed-dotted line).}
\label{fig5}
\end{figure*}

The exciton plus electron structure of the excited states show very
similar tendency. Fig. \ref{fig5} shows the correlation functions for
the 4 resonance states. $E_1$ and $E_2$ are below the $2s$ exciton
threshold and they have a pronounced $2s$ exciton plus an outer electron
structure. The electron-hole correlation function of the excited state
trion is very similar to the electron-hole density in the exciton. The difference
between the excited trion with energy $E_1$ and $E_2$ is that in the 
latter the second electron is much farther away from the exciton (as
the electron-electron correlation function shows in Fig. \ref{fig5}). 
In the trion with  energy $E_2$ the electron-hole correlation has a long tail overlapping 
with the electron electron-correlation function. The second electron
strongly polarizes the exciton. The excited trion binding is due to the
significant overlap between the electron-electron
and electron-hole correlation functions.  The excited states with energy
$E_3$ and $E_4$ show very similar tendency, an $3s$ exciton plus an
electron far out from the center.

The energies and the average distances between particles are compared
in Table I for different $\tilde{r}_0$ values. Compared to the compact
ground state trions, the electron-hole and electron-electron distances
are very large in the  excited states. The large size is explained by the facts that the
excited exciton is larger and the are loosely bound states are more
extended. Table I. also shows 
\begin{equation}
\delta_{ee}=C_{ee}(0), \ \ \ {\rm and} \ \ \ \ 
\delta_{eh}=C_{eh}(0) 
\end{equation}
the probability that the two electron or an electron and the hole are
at the same spatial position. This probability is decreasing with the
increased spatial distribution and the probability for electron-hole
is always larger than that of the electron-electron.

To calculate  resonance states for $0<\tilde{r}_0$ we use the 
ECG basis that is  generated by SVM for $\tilde{r}=0$. The basis
dimension is very large $(K=1500)$ and the basis is flexible enough 
to be accurate for nonzero $\tilde{r_0}$. Actually the $\tilde{r}_0=0$
case is the most challenging calculation because of the deep Coulomb
potential near the origin. Increasing $\tilde{r}_0$ leads to less
attractive potential close to the origin, while the asymptotic part
($r>5$) of the potential still behaves like $1/r$. This asymptotic Coulomb
part determines the resonance wave functions beyond $r=5$ so one expects
that the resonances also exist for $0<\tilde{r}_0$. 

\begin{figure*}
\subfigure{
\includegraphics[width=0.45\linewidth,clip]{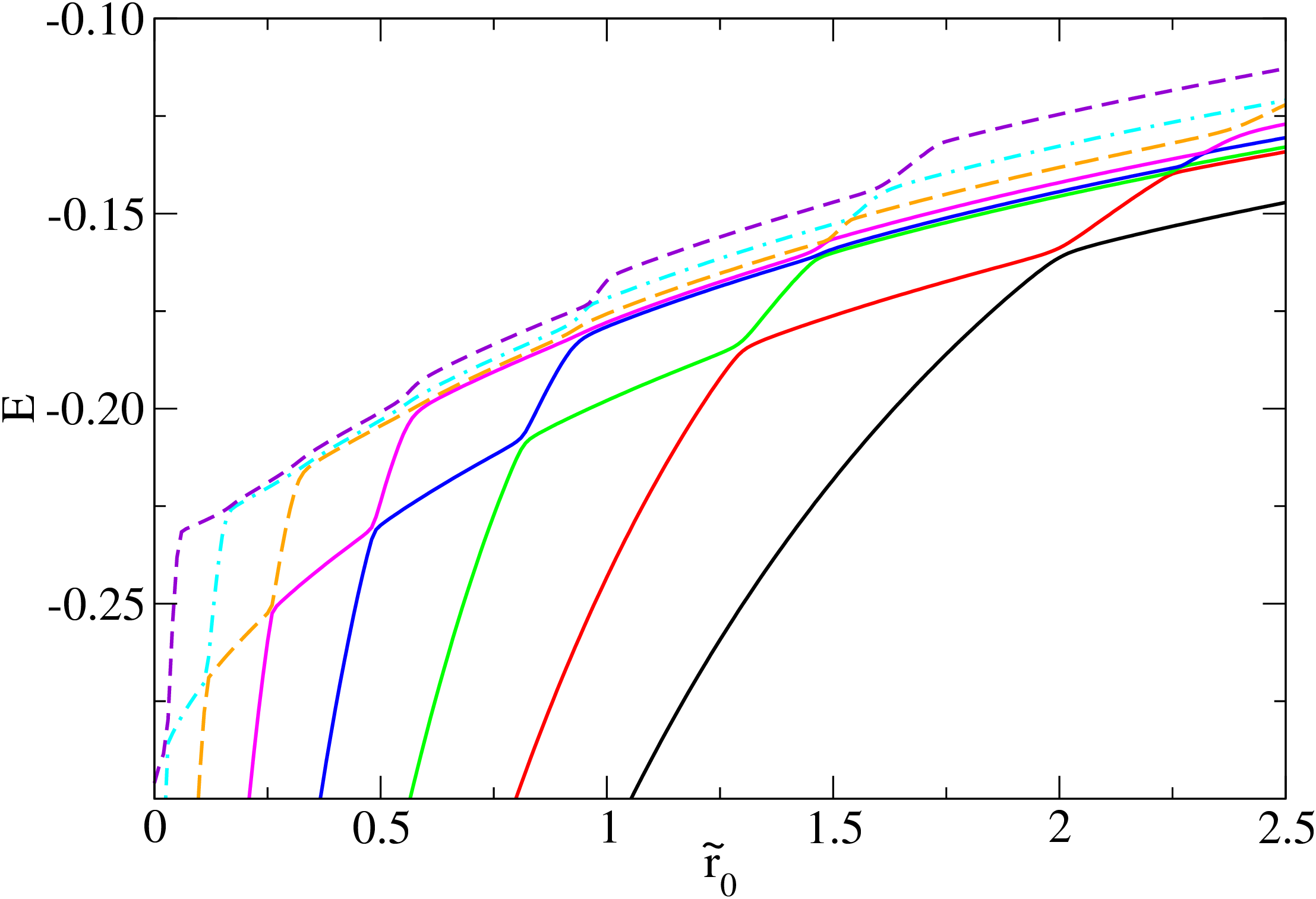} 
}
\subfigure{
\includegraphics[width=0.45\linewidth,clip]{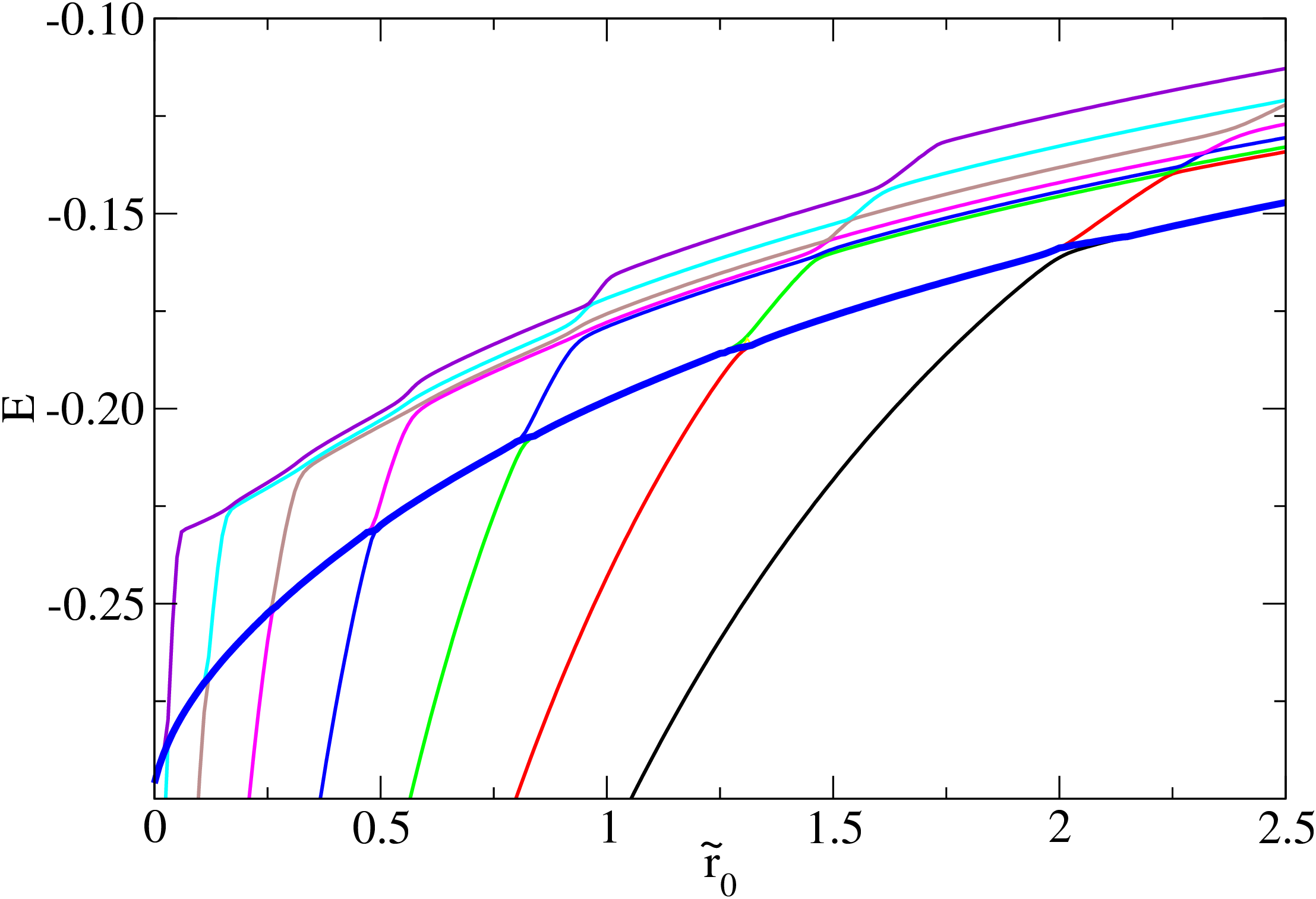}
}
\caption{Change of energy levels as a function of $\tilde{r}_0$. Left:
By increasing $\tilde{r}_0$ the potential weakens and the energy levels move
upward. The energy of $E_1$ moves upward but at around $\tilde{r}_0$=0.02
the energy of the state below $E_1$  moves up faster (dot dashed line) 
becomes nearly equal the energy of $E_1$. The two energy level cannot cross
each other (avoided crossing) and $E_1$ continues on the dot dashed line.
The same thing happens at $\tilde{r}_0$=0.12, and $E_1$ continues on the long
dashed line, and so on. The path of $E_1$ as a function of $\tilde{r}_0$ is
shown in the right (thick line).
}
\label{fig6}
\end{figure*}

Using the ECG basis optimized for $\tilde{r}_0=0$ we show the change of
the calculated resonance energies as a function $\tilde{r}_0$ in Fig. \ref{fig6}.
We have optimized the ECG basis for several $\tilde{r}_0$ values and
used CS to check that the resonance energy trajectory in Fig. \ref{fig6}
is accurate. 

\begin{figure*}
\subfigure{
\includegraphics[width=0.45\linewidth,clip]{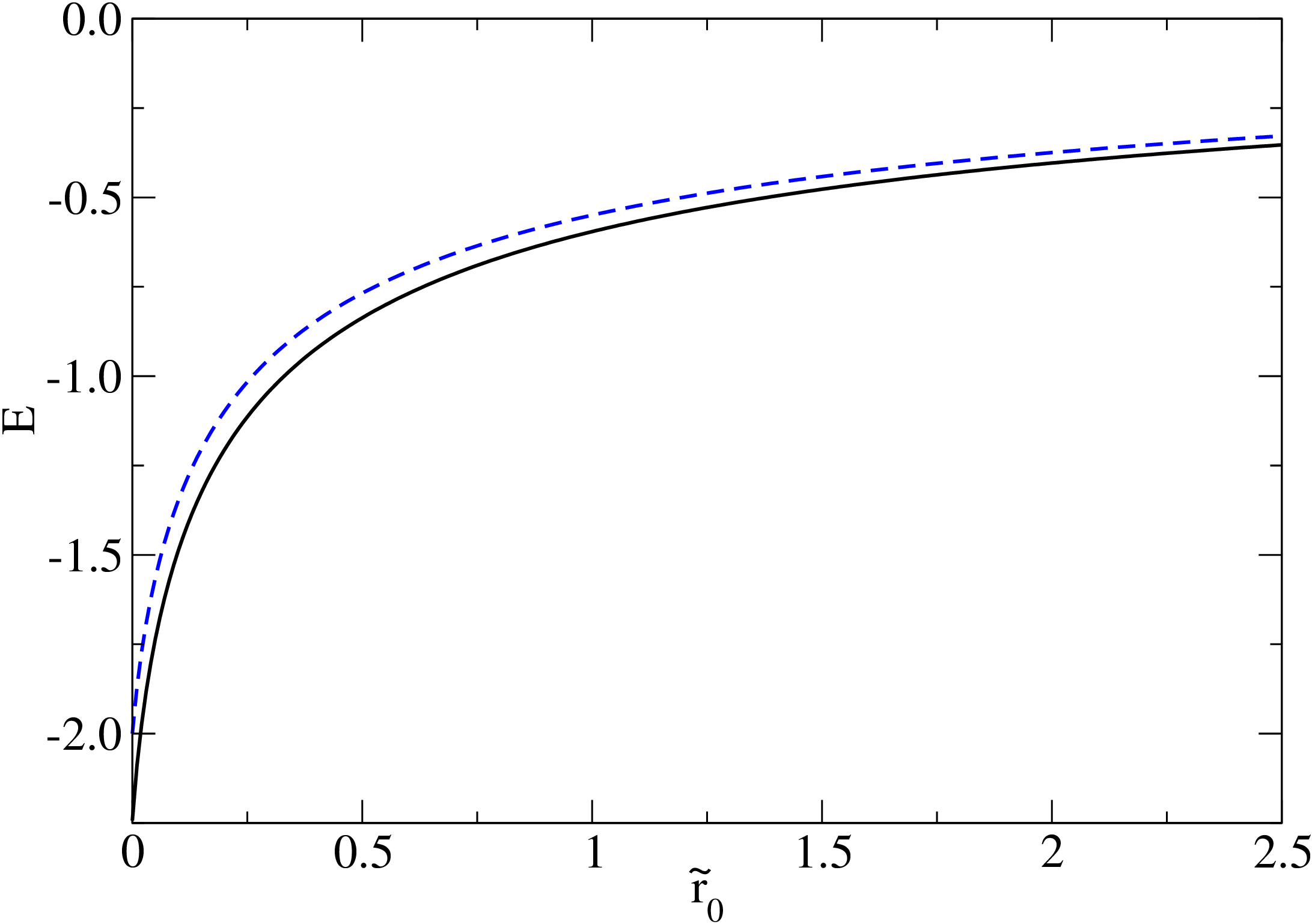} 
}
\subfigure{
\includegraphics[width=0.45\linewidth,clip]{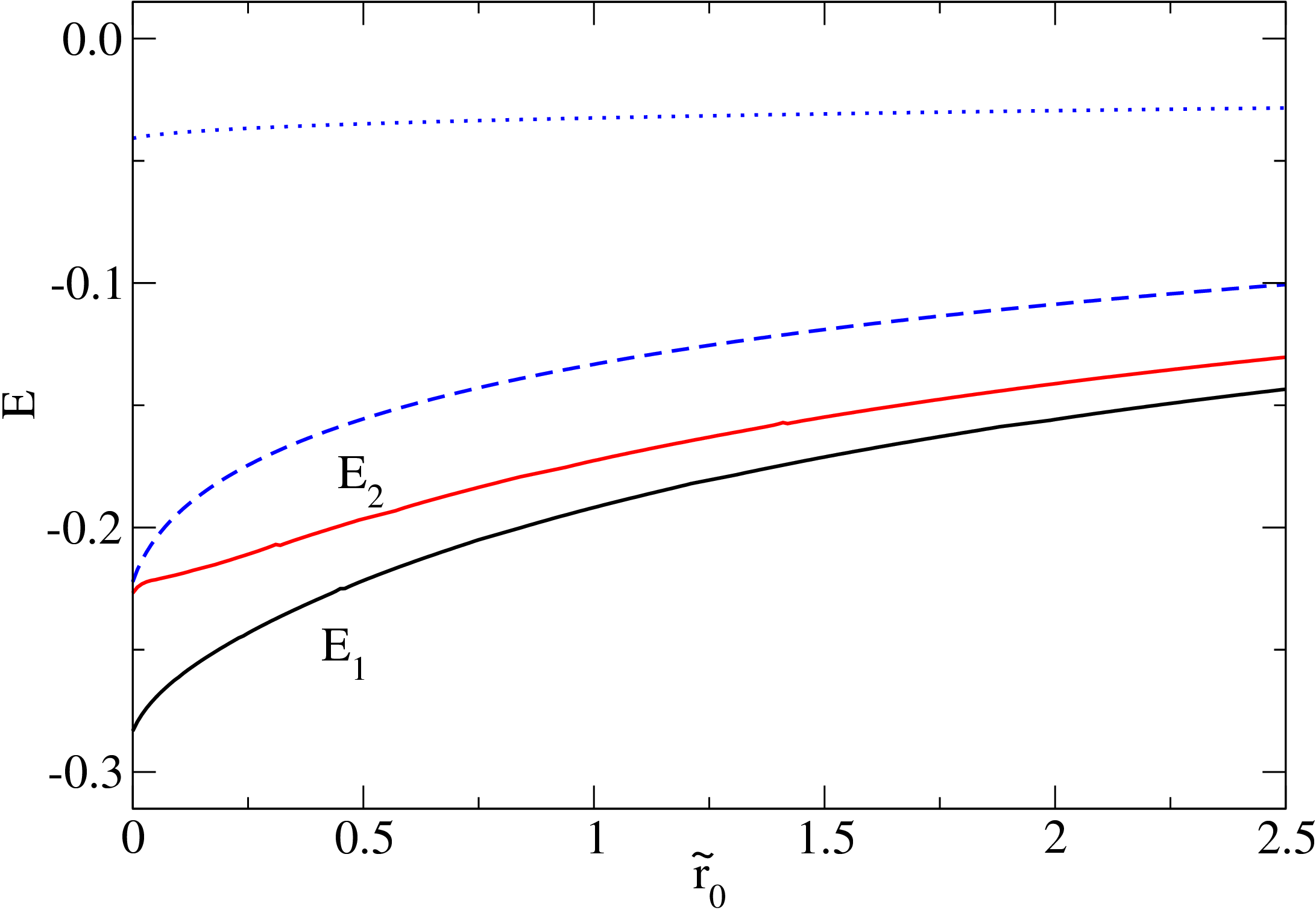}
}
\caption{Energies as a function of $\tilde{r}_0$. Ground state energy trion (solid
line), $1s$ exciton energy (dashed like) (left); Excited state
energies of trion (solid line), $2s$ exciton energy (dashed line), $3s$ exciton energy
(dotted line) (right).
}
\label{fig7}
\end{figure*}

The calculated total energies and the binding energies for the ground
state and the lowest two resonance states are shown in Figs.
\ref{fig7} and \ref{fig8}. We only show these two resonance states because
the energy of the higher states barely depends on $\tilde{r}_0$ (see.
Table I.). This is not surprising, the resonance states with higher
energy only feel the $1/r$ tail of the potential which is independent
of $\tilde{r}_0$. The weak $\tilde{r}_0$ dependence is also  true for
the resonances with energy $E_1$ and $E_2$ above $\tilde{r}_0=1$. The 
mean distances between particles at $\tilde{r}_0=1$ are larger
than $r=5$ (see Table I.) so these states are mostly affected by the Coulomb tail. 
Energies $E_1$ and $E_2$ proportional to the energy of the $2s$ trion
(see Fig. \ref{fig7}) and the binding energy hardly changes (see Fig.
\ref{fig8}). 

The most interesting feature of the dependence of the
binding energies  on $\tilde{r}_0$ is that the excited states become more
bound than the ground state by increasing $\tilde{r}_0$. The
trajectory of $E_1$ and $E_0$ crosses at $\tilde{r}_0=0.55$ and
$E_2$ and $E_0$ crosses at $\tilde{r}_0=1.4$.  This happens because
the ground state wave function (see Fig. \ref{fig4}) is nonzero close
to the origin and increasing $\tilde{r}_0$ weakens the potential in that
region and the ground state binding energy rapidly decreases with $r_{0}$ as
Fig. \ref{fig8} shows. As it was already mentioned, the excited states
are mostly governed by the Coulomb tail and their energies are less
sensitive to $\tilde{r}_0$.

The correlation functions of the excited states for $\tilde{r}_0>0$ 
is very similar to the $\tilde{r}=0$ case shown in Fig. \ref{fig5}, 
but spatially more extended. This is because increasing $\tilde{r}_0$ the total 
energies are decreasing and the size of the states are increasing
(see Table I.). As an illustration for the similarity we show the 
ground state trion for $\tilde{r}_0=0$ and $\tilde{r}_0=1$
in Fig. \ref{fig4}. 

We have also investigated the effect of electron and hole mass ratio
on the binding energy. By using different masses the electron hole
the electron hole symmetry is broken so we have two different trions, 
$ehh$ and $eeh$. To magnify the binding energy differences, we have
used a relatively large hole mass by choosing $\sigma=2/3$. For for
smaller hole masses the tendencies in the binding energies can not be
easily illustrated. Fig. \ref{fig9} shows the binding energy of $ehh$
and $eeh$ as a function of $\tilde{r}_0$. The heavier hole leads to
larger binding energies compared to the $\sigma=1$ case, and the $ehh$
system is has larger binding energy than $eeh$, except for the $E_2$
resonance where the two energies are nearly equal. The mass difference 
also affects the crossing points.

\subsection{Spin triplet case}
Similarly to the 3D case \cite{PhysRevA.66.010502}, we have also found
two resonances for  spin triplet trions in 2D, one right below the
$2s$ threshold and one very close to the $3s$ threshold, so the binding
energies of these resonances is very small (see Table II.). The average 
distances in these systems are very similar to the those of the singlet 
$E_1$ and $E_3$, respectively.
\begin{widetext}
\begin{table*}
\begin{tabular}{|c|l|l|l|l|l|l|l|}
\hline
$E_r$   & $E_i$                & $r_{eh}$&$r_{eh}^2$&$\delta_{eh}$& $r_{ee}$&$r_{ee}^2$&$\delta_{ee}$   \\\hline
-0.2254 & -5.0$\times 10^{-6}$ &11.94   &256.66    &0.16     &22.88& 599.55   & 1.5$\times 10^{-12}$ \\
-0.0845 & -1.2$\times 10^{-6}$ &19.61   &573.79    &0.03     &37.72& 1568.04  & 1.9$\times 10^{-11}$\\
\hline
\end{tabular}
\caption{Energies and other properties of trion states for $S=1$,
$\tilde{r}_0$=0 and $\sigma=1$.}
\end{table*}
\end{widetext}
Fig. \ref{fig10} compares the correlation function of the exciton
and the triplet  trion resonances. Once again, the
electron-hole correlation function is very similar to the $2s$
and $3s$ exciton. The peak of electron-electron correlation function,
however, is almost twice as far away then in the singlet case 
(see Fig. \ref{fig5} $E_1$ and $E_3$) because the two electron has parallel spins.

\subsection{Comparison to experiments}
In this section we compare the calculated results to experimental
measurements in monolayer TMDs sandwiched between hexagonal boron nitrade.
The encapsulation of TMD monolayers between atomically smooth hexagonal 
boron nitrade layers allow high quality optical measurements. Binding
energies and radii of Rydberg exciton states and energies of charged
excitons were measured in WS$_2$
\cite{PhysRevLett.123.167401,PhysRevLett.113.076802}, WSe$_2$
\cite{PhysRevLett.113.026803,PhysRevB.96.085302,Lyons2019,PhysRevLett.123.027401,
Jones2016,Ye2018}, MoS$_2$
\cite{Roch2019,Goryca2019,trion_MoS2,Mak2013,PhysRevLett.105.136805,PhysRevMaterials.2.011001},
MoSe$_2$ \cite{Goryca2019,PhysRevX.8.031073}
and MoTe$_2$ \cite{Goryca2019,PhysRevX.8.031073}. 

Using the  parameters ($\mu,r_0$, and $\kappa$) given in  Table I.
of Ref. \cite{Goryca2019} we have calculated the binding energies and root
mean square distances of excitons and trions in WS$_2$, WSe$_2$, MoSe$_2$, MoS$_2$ and 
MoTe$_2$ (Table III.). Note, that in some cases the $\tilde{r}_0$ values are very different, but 
the energies are very similar. For example, 
the binding energies of WS$_2$ and WSe$_2$ are very close despite of
the difference in $\tilde{r}_0$. This is because the change of binding energy between
$\tilde{r}_0$=0.59 and $\tilde{r}_0$=0.84 is compensated by the
slightly larger $E^*$ of WSe$_2$. $E^*$ is inversely proportional to $\kappa^2$ so small
changes in $\kappa$ can cause large energy changes.

The calculated binding energies and radii for 
excitons (Table III.) are in excellent agreement with the experimental binding energy and
radii of Table I of \cite{Goryca2019}, reproducing the fit of the model to experimental
data of Ref. \cite{Goryca2019}. The calculated exciton radii and energies are 
also in agreement with the calculated and the experimental values 
for WSe$_2$
\cite{prl.120.057405} 
(the calculated $r^2_{eh}$=1.67 nm for $1s$ and $r^2_{eh}$=6.96 nm for
$2s$, the experimental values are 1.7 nm and 6.6 nm, the  calculated  $E_{2s}-E_{1s}$ is 124 meV, 
the experimental value is 130 meV). 

In previous calculations (see Table II. of Ref.
\cite{PhysRevB.93.125423}) for TMDs suspended in vacuum 
or placed on SiO$_2$ substrate 
the calculated and experimental energies of excitons were 50-100 meV different.
The new  and more accurate measurements using monolayer TMDs sandwiched between hexagonal boron nitrades
allow the study of the Rydberg states of excitons in magnetic field and one can extract the binding energy
and radii of the Rydberg states \cite{Goryca2019}. These physical properties then can be used to find
the most suitable model parameters, reducing the difference between the experimental and theoretical binding 
energy of excitons to less then 5 meV. 

The agreement of calculated and experimental binding energies (Table III.) are not as good
as for excitons, but in general it is similar to the agreement for the TMDs suspended in vacuum
or placed on SiO$_2$ surface \cite{PhysRevB.93.125423,PhysRevB.96.035131}. Comparing the experimental
and calculated trion energies one has to keep in mind that the model parameters were fitted to the exciton
measurements \cite{Goryca2019}, but the trion energies was measured in
the same experiment. For example, the $E_{2s}-E_{1s}$
energy difference was measured to be 141.7 meV for WS$_2$ in Ref.
\cite{Goryca2019}. The value of
this transition energy is very important to find the screening length.
In Ref. \cite{PhysRevLett.123.167401} trion states at 31 and 37 meV were reported 
in WS$_2$, but
the $E_{2s}-E_{1s}$ difference was measured to be 145 meV in the this experiment.  
This small difference (141.7 meV
to fit the parameter and 145 meV in the measurement) would 
lead to relatively large change in the model parameters and would affect
the binding energy of the trions. 
Another factor to consider is the effective mass, $\mu$.
The energy of the exciton only depends on $\mu$ so the experiments can only pinpoint the reduced mass but give no guidance
about effective mass of the hole. In the calculations presented in Table III. we used $m_e=m_h$ ($\sigma=1$).
As it is illustrated in Fig. \ref{fig9}, using different values for the electron and hole mass while
keeping the reduced mass the same increases the binding energies. For
the energy of trions in TMDs (see Table III.)
this would lead to a 3-5 meV increase in binding energies. 

The calculated binding energy for the excited-state trions is about the same range as the ground
state binding energies, these excited states have relatively large binding energies. Excited-state trion was recently 
reported in WS$_2$ \cite{PhysRevLett.123.167401} and MoSe$_2$ \cite{yan} with binding energies close
to the binding energy of the ground state. The width of these
resonance states are very small and these are 
quasi-bound states. In the lower excited state, $E_1$, the $eh$ and
$ee$ distances about 2.5 to 3 times
larger than in the ground state. This state is an $2s$ exciton with an loosely bound electron (or hole) circling 
around it. The second excited state, $E_2$, is even larger with tens of nanometers of distances  between $ee$ and $eh$. 
Due the large size this state is a model prediction in a perfect 2D system, but it is unlikely that this can be
measured in a real material.

\begin{figure*}[htp]
\centering
\includegraphics[width=0.95\linewidth]{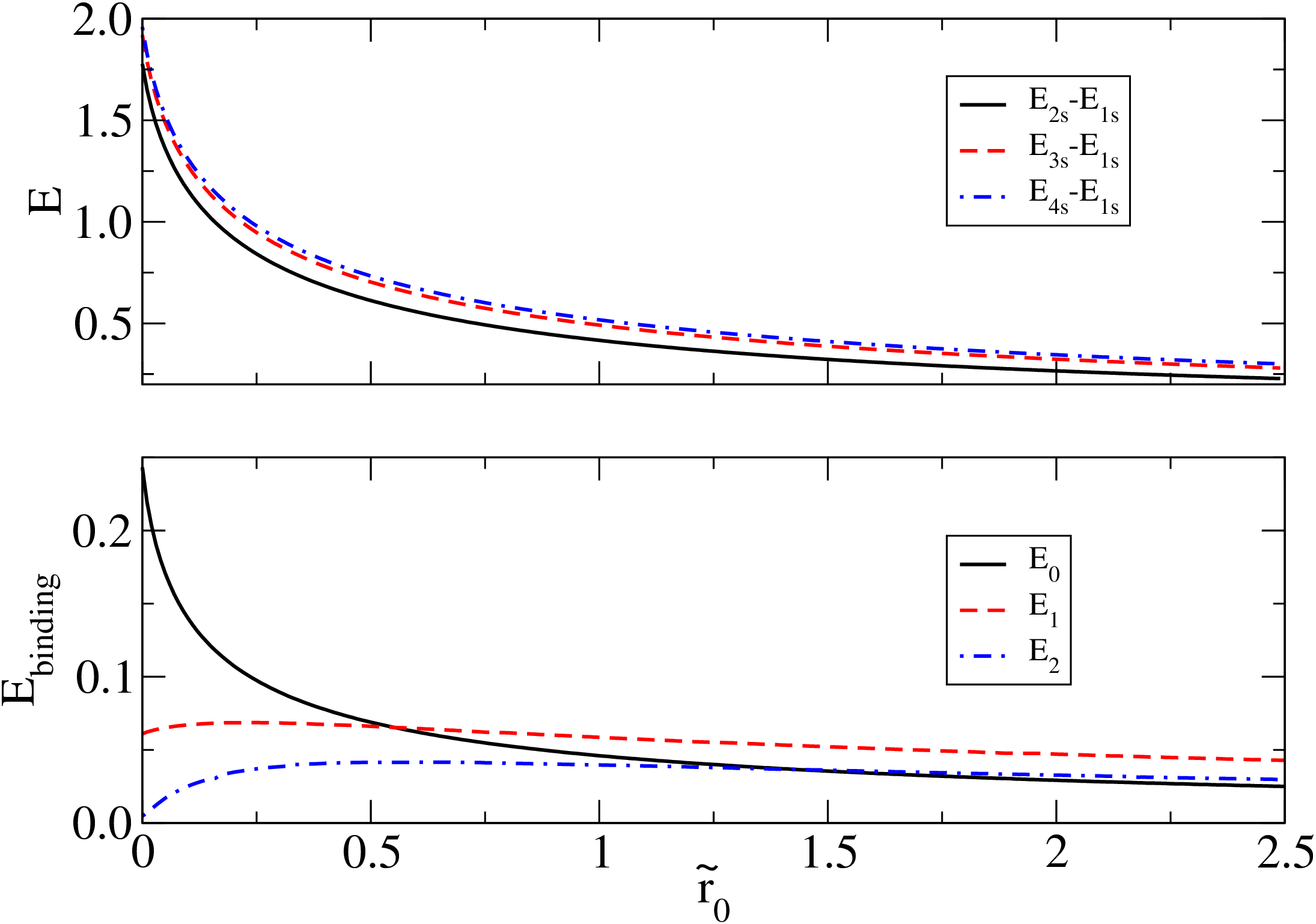} 
\caption{Top: Transition energies of exciton excited states as a function of $\tilde{r}_0$.
Bottom: Binding energies of the ground and the two lowest resonance
states the ground and the two lowest resonance 
states as a function of $\tilde{r}_0$.}
\label{fig8}
\end{figure*}

\begin{figure*}[htp]
\centering
\includegraphics[width=0.95\linewidth]{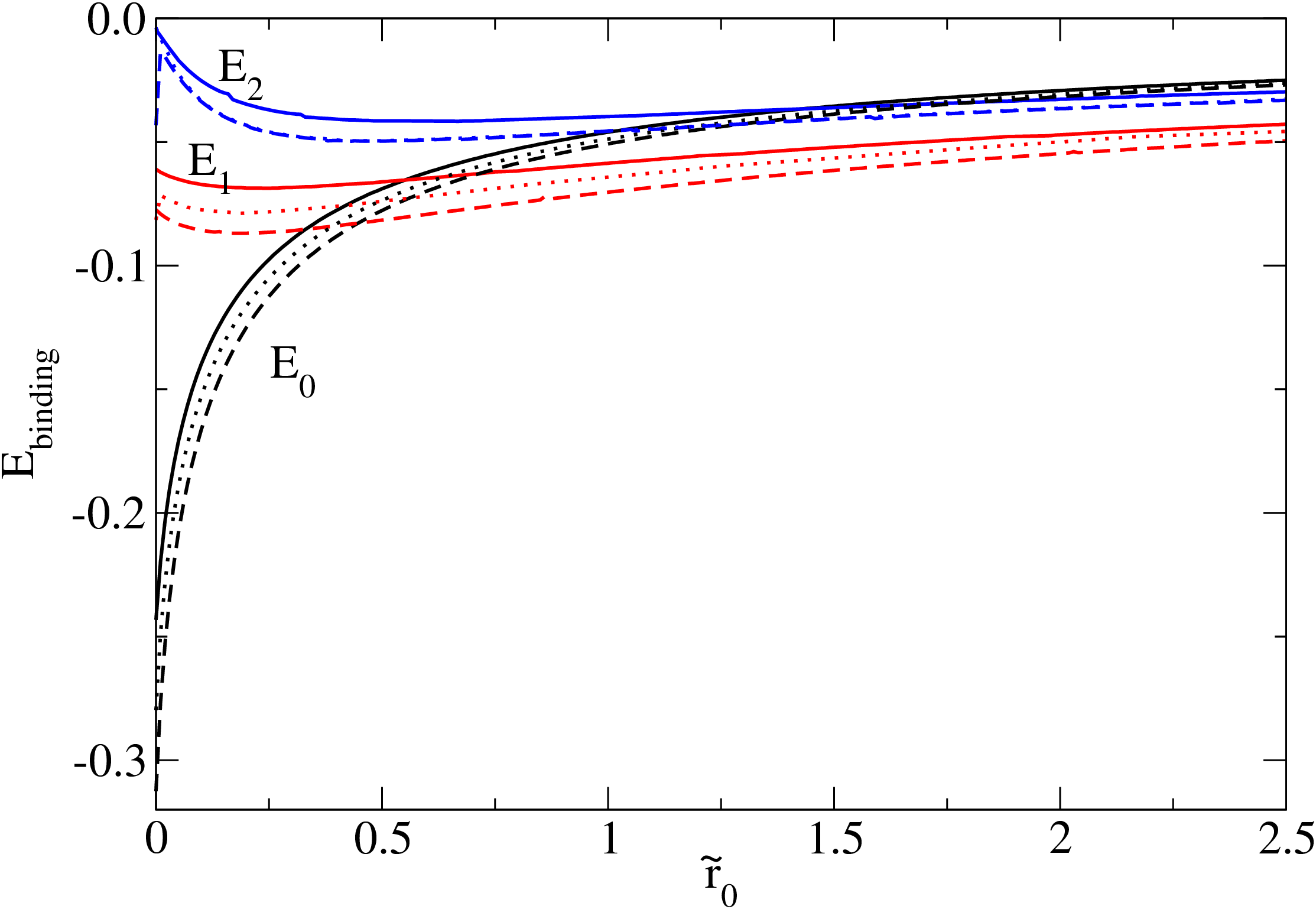} 
\caption{Binding energies as a function of $\tilde{r}_0$ (solid line 
$eeh$ with $\sigma=1$, dotted line $eeh$ with $\sigma=2/3$, dashed line
$ehh$ with $\sigma=2/3$.}
\label{fig9}
\end{figure*}

\begin{figure*}[htp]
\centering
\includegraphics[width=0.95\linewidth]{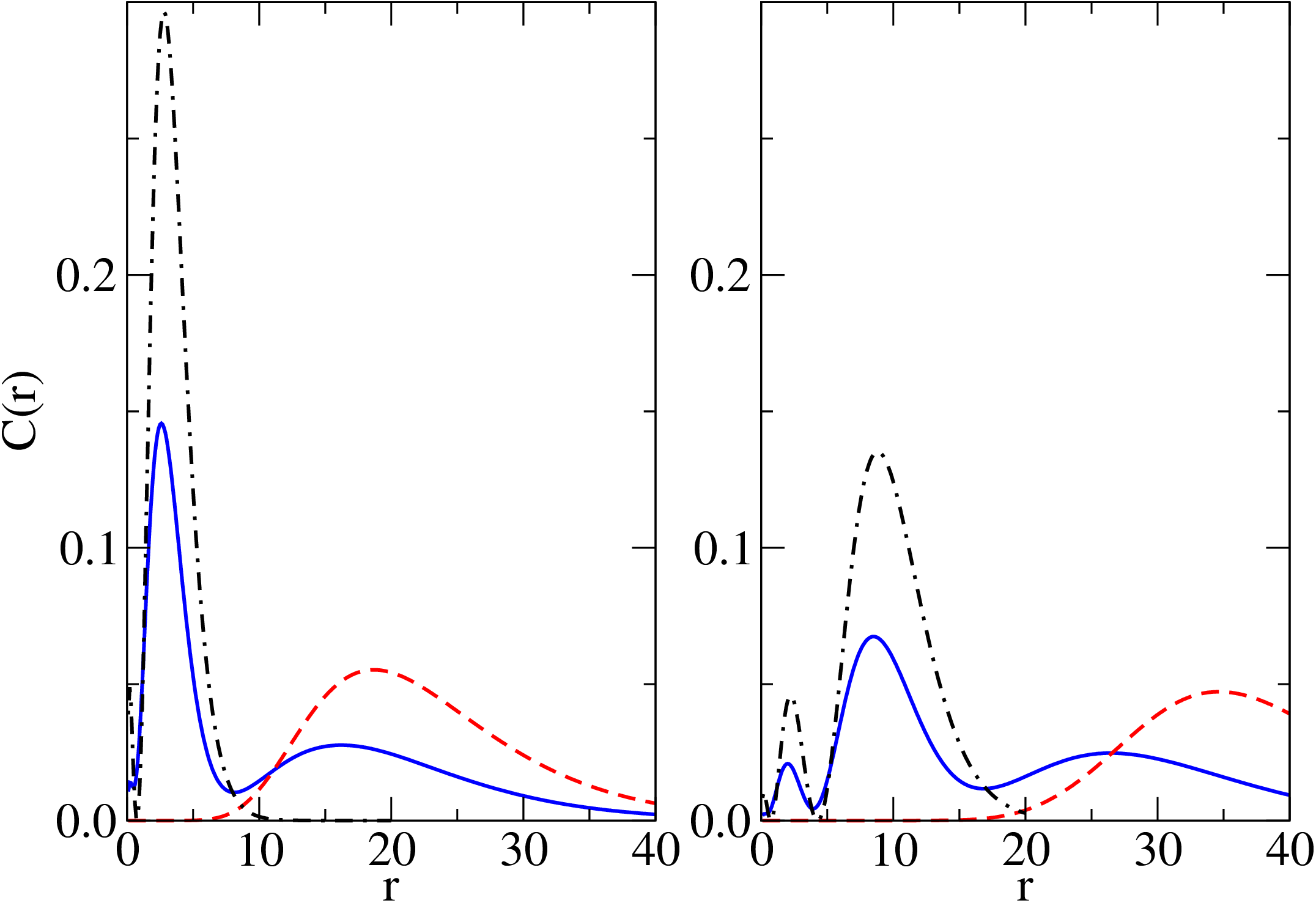} 
\caption{Ground state correlation functions for $S=1$ resonances
below the $2s$ threshold (left) and below the $3s$ threshold (right).
Electron-electron correlation, $C^t_{ee}(r)$ (dashed line),
electron-hole correlation, $C^t_{eh}(r)$ (solid line), and 
exciton electron-hole correlation, $C^x_{ee}(r)$ (dashed-dotted line for
$2s$ exciton (left) and $3s$ exciton (right).}
\label{fig10}
\end{figure*}

Fig. \ref{fig8} can be used as a guide to analyze the agreement between
the Keldysh potential based models and experiments. If the energy
differences between the Rydberg exciton states are measured, one
can choose the most suitable $\tilde{r}_0$ value using the top part of 
Fig. \ref{fig8}. This value then can be used to predict the trion
energies using the bottom part of Fig. \ref{fig8}.

\section{Summary}
Using the complex scaling and the stabilization method combined with the stochastic variational approach, we
have studied resonance states of three-particle systems interacting with a Coulomb and a screened Coulomb
(Keldysh potential). The stochastic variational method was used to generate a suitable square integrable basis
of explicitly correlated Gaussians. The stochastic variational method has been previously used
\cite{PhysRevB.93.125423,PhysRevB.96.035131} to describe trions, biexcitons and charged biexcitons in 
TMDs and in high precision calculations in atomic and molecular systems \cite{appliedECG}. We have used two
independent approaches, the complex scaling and the stabilization to calculate the resonance states using real basis
functions.

In 2D Coulomb three particles systems with $S=0$, we have found three resonance states below the $2s$ and two resonance states 
below the $3s$ two-particle (exciton) threshold. These states can be envisioned as a $2s$ or $3s$ exciton 
with a third, loosely bound particle circling  around it. Comparing the correlation functions of $2s$ and $3s$ 
excitons to those of the excited trions confirm that picture. We have also found 2 resonance states for $S=1$. Resonance
states similar to these have been studied in 3D for Coulomb potential 
\cite{PhysRevA.66.010502}. These resonances are mostly due to 
the long tail of the Coulomb interaction and they survive the confinement 
from 3D to 2D, despite of the fact that their
energy and spatial extension radically changes. 

Screening the Coulomb interaction using a Keldysh potential changes the energy of these resonances, but they remain narrow quasi-bound states. The screening decreases the Coulomb potential at the origin, pushing the resonance wave function
farther out, so the interparticle distances quickly increases. As the wave functions of the resonance states are
mostly feel the Coulomb tail, the resonance energies are less sensitive to the screening than the energy of
the ground state. At some screening length the binding energy of the
resonance states will be larger than that of
the ground state.

We have calculated the energies and interparticle distances of these resonance states for various TMDs.
Energies of excited-state trions in WS$_2$ \cite{PhysRevLett.123.167401}
and MoSe$_2$ \cite{yan} are reasonably close to the values predicted by the calculations. Closer agreement
may require more elaborate calculations including spin-orbit interactions and multiband Hamiltonians as used, for example, 
in Ref. \cite{PhysRevB.96.035131}.

\begin{acknowledgments}
K.V. was supported by the National Science
Foundation (NSF) under Grant No. IRES 1826917,
J.Y. acknowledges support by NSF ECCS-1509599. 
\end{acknowledgments}

\begin{widetext}
\begin{table*}
\begin{tabular}{|c|l|l|l|l|l|l|l|}
\hline
system& property       &method &WS$_2$   &WSe$_2$          &MoS$_2$      &MoSe$_2$   &MoTe$_2$ \\\hline
  & $\tilde{r}_0$ &            &0.5942   &0.8399           &0.8923       &1.3324     &2.2489   \\\hline
  & $ E_{1s} $    &calc        &-178.7   &-161.4           &-220.2       &-231.9     &-176.9   \\\cline{2-8}
  & $ E_{1s} $    &exp         &-180$^a$ &-167$^a$,-170$^h$&-221$^{a,i}$ &-231$^a$   &-177$^a$  \\\cline{2-8}
  & $ \sqrt{<r^2_{eh}>}$       &calc     &1.66             &  1.68       & 1.23      & 1.10      & 1.31    \\\cline{2-8}
$eh$ & $ \sqrt{<r^2_{eh}>}$    &exp      &1.8$^a$          &  1.7$^{a,b}$& 1.2$^a$   & 1.1$^a$   & 1.3$^a$ \\\cline{2-8}
  & $ E_{2s} $    &calc        &-37.8    & -37.4           & -51.8       &  -60.6    &-52.8       \\\cline{2-8}
  & $ \sqrt{<r^2_{eh}>}$       &calc     &7.20             &  6.97       & 5.08      &  4.32     & 4.81    \\\cline{2-8}
  & $ \sqrt{<r^2_{eh}>}$       &exp      &                 &  6.6$^b$    &           &           &         \\\cline{2-8}
  & $E_{2s}-E_{1s}$&calc       &140.9    & 124             & 168.4       &171.3      &124.1        \\\cline{2-8}
  & $E_{2s}-E_{1s}$&exp        &141.7$^a$&130$^b$,131$^h$  & 170$^{a,i}$ &168$^{a}$,148$^j$,152$^c$ &124$^{a,j}$  \\\hline
 \hline
  & $ E_0     $    &calc   &-194.4   & -175.2      &-238.8   &-250.8     &-190.5   \\ \cline{2-8}
  & $ E_0^b   $    &calc   & 15.7    &  13.8       &  16.6   & 18.9      & 13.6    \\ \cline{2-8}
  & $ E_0^b   $    &exp    &27,31$^d$&21,29$^e$    &17,25$^f$&27$^{c,g}$ &         \\ \cline{2-8}
  & $ \sqrt{<r^2_{eh}>}$    &calc   & 3.01    &  3.05       &  2.2    & 2.0       & 2.4     \\ \cline{2-8}
  & $ \sqrt{<r^2_{ee}>}$    &calc   & 4.24    &  4.03       &  3.2    & 2.8       & 3.4     \\\cline{2-8}
  \cline{2-8}
  & $ E_1     $    &calc   &-54.1    & -53.8       & -74.4   & -87.3     &-75.5    \\ \cline{2-8}
  & $ E_1^b   $    &calc   & 16.3    &  16.4       &  22.6   &  26.6     & 22.7    \\ \cline{2-8}
$eeh$ & $ E_1^b$   &exp    & 22$^d$  &             &        & 27$^c$    &         \\ \cline{2-8}
  & $ \sqrt{<r^2_{eh}>}$    &calc   &  8.5    &  8.83       &  6.1    &  5.5      & 5.8     \\ \cline{2-8}
  & $ \sqrt{<r^2_{ee}>}$    &calc   & 13.9    & 14.20       &  9.9    &  8.8      & 9.4     \\\cline{2-8}
  \cline{2-8}
  & $ E_2     $    &calc   & -48.0   & -48.1       & -66.8   & -78.8     &-67.9    \\ \cline{2-8}
  & $ E_2^b   $    &calc   &  10.2   &  10.7       &  15.1   &  18.2     & 15.1    \\ \cline{2-8}
  & $\sqrt{<r^2_{eh}>}$    &calc   & 54.9    &  45.8       &  29.0   &  28.2     & 29.3    \\ \cline{2-8}
  & $\sqrt{<r^2_{ee}>}$    &calc   & 77.6    &  64.9       &  41.1   &  40.0     & 41.4    \\\hline
\end{tabular}
\caption{Energies and mean distances in TMDs. Energies are in meV,
distances and in nm, $E_1^b=E_1-E_{1s}, 
E_2^b=E_2-E_{2s}, E_3^b=E_3-E_{2s}$.
The experimental values are taken from:
$^a$ \cite{Goryca2019},
$^b$ \cite{prl.120.057405},
$^c$ \cite{yan},
$^d$ \cite{PhysRevLett.123.167401},
$^e$ \cite{Chen2018},
$^f$ \cite{Roch2019},
$^g$ \cite{PhysRevLett.123.027401}
$^h$ \cite{doi:10.1021/acs.nanolett.9b00029},
$^i$ \cite{PhysRevMaterials.2.011001},
$^j$ \cite{PhysRevX.8.031073}
}
\end{table*}
\end{widetext}

%

\end{document}